 \documentclass[final,3p,10pt,times]{elsarticle}

\usepackage{amssymb}
\usepackage{amsmath}
\usepackage{ntheorem}
\usepackage{amssymb}
\usepackage{hyperref}

\usepackage{caption}
\usepackage{tikz}
\usepackage{xcolor}
\usepackage{hyperref}
\usepackage{graphicx}
\usepackage{subcaption}

\usepackage{setspace}
\usepackage{epstopdf}
\usepackage{enumitem}
\usepackage{booktabs} 
\usepackage{multirow} 
\usepackage{booktabs}
\usepackage{makecell}
\usepackage{float}

\usepackage{booktabs}
\usepackage{multirow}

\makeatletter

\newcommand{\Rmnum}[1]{\expandafter\@slowromancap\romannumeral #1@}
\makeatother

\journal{Digital Signal Processing}

\begin{document}
\begin{sloppypar}
	
\begin{frontmatter}



\title{Graph Fractional Hilbert Transform: Theory and Application}

 \author[1]{Daxiang Li}
 \author[1,2,3,4]{Zhichao Zhang\corref{cor1}}

\ead{zzc910731@163.com}
\cortext[cor1]{Corresponding author; Tel: +86-13376073017.}
\address[1]{School of Mathematics and Statistics, Nanjing University of Information Science and Technology, Nanjing 210044, China}
\address[2]{Hubei Key Laboratory of Applied Mathematics, Hubei University, Wuhan 430062, China}
\address[3]{Key Laboratory of System Control and Information Processing, Ministry of Education, Shanghai Jiao Tong University, Shanghai 200240, China}
\address[4]{Key Laboratory of Computational Science and Application of Hainan Province, Hainan Normal University, Haikou 571158, China}

\tnotetext[mytitlenote]{This work was supported in part by the Open Foundation of Hubei Key Laboratory of Applied Mathematics (Hubei University) under Grant HBAM202404; in part by the Foundation of Key Laboratory of System Control and Information Processing, Ministry of Education under Grant Scip20240121; and in part by the Foundation of Key Laboratory of Computational Science and Application of Hainan Province under Grant JSKX202401.}


%

\begin{abstract}
The graph Hilbert transform (GHT) is a key tool in constructing analytic signals and extracting envelope and phase information in graph signal processing. However, its utility is limited by confinement to the graph Fourier domain, a fixed phase shift, information loss for real-valued spectral components, and the absence of tunable parameters. The graph fractional Fourier transform introduces domain flexibility through a fractional order parameter $\alpha$ but does not resolve the issues of phase rigidity and information loss. Inspired by the dual-parameter fractional Hilbert transform (FRHT) in classical signal processing, we propose the graph FRHT (GFRHT). The GFRHT incorporates a dual-parameter framework: the fractional order $\alpha$ enables analysis across arbitrary fractional domains, interpolating between vertex and spectral spaces, while the angle parameter $\beta$ provides adjustable phase shifts and a non-zero real-valued response ($\cos\beta$) for real eigenvalues, thereby eliminating information loss. We formally define the GFRHT, establish its core properties, and design a method for graph analytic signal construction, enabling precise envelope extraction and demodulation. Experiments on edge detection, anomaly identification, and speech classification demonstrate that GFRHT outperforms GHT, offering greater flexibility and superior performance in graph signal processing.

\end{abstract}




\end{frontmatter}
\section{Introduction}
Graph signal processing (GSP) has established itself as a fundamental paradigm for the analysis and processing of data residing on irregular structures, finding widespread applications in social networks \cite{1,2,3,4,5,6,7,8}, sensor arrays \cite{7,9,10,11,12,13,14}, brain connectomics \cite{15,16,17,18,19,20,21,22,23}, and power grids \cite{4,13,24,25,26,27,28}. A pivotal task in GSP transcends mere vertex-domain processing; it involves the precise extraction of spectral and phase characteristics of graph signals \cite{1,4,6,29,30,31,32,33,34,35,36}. This capability is crucial for advanced applications such as signal demodulation, envelope detection, and the identification of transient features or anomalies within complex network data \cite{37,38,39}.

To facilitate this analysis, the graph Hilbert transform (GHT) is introduced as a direct analogue of its classical counterpart within the GSP framework \cite{37}. The GHT extracts envelope and instantaneous phase information by constructing a graph analytic signal (GAS), and is consequently widely used in tasks such as image edge detection, anomaly detection, and speech processing \cite{38}.

Despite its demonstrated utility, the conventional GHT is constrained by several inherent and structural limitations that impede its effectiveness in processing modern, complex graph signals:
\begin{enumerate}[label=(\roman*)]
    \item \textbf{Confinement to a single spectral domain:} The GHT operates exclusively within the canonical graph Fourier transform (GFT) domain \cite{6,31,40,41,42,43}. This restricts analysis to a single perspective, lacking a mechanism for smooth transition between the vertex and spectral domains. Consequently, it fails to capture hybrid vertex-spectral features that are prevalent in dynamic and complex networks.
    \item \textbf{Fixed and inflexible phase shift:} The transfer function of the GHT is restricted to a small set of discrete values $\{\pm j, 0\}$. This rigidity prevents any continuous tuning of the phase shift, which is often essential for precise phase manipulation and tailored analysis.
    \item \textbf{Inherent information loss:} A critical shortcoming of the GHT is its assignment of a zero response to all real-valued eigenvalues ($\operatorname{Im}(\lambda_k)=0$). For signals defined on undirected graphs or graphs with real spectra, this operation irrevocably discards associated signal components, leading to a loss of energy and potentially meaningful information.
    \item \textbf{Lack of adaptability:} The GHT is a static, non-parameterized transform. It possesses no tunable parameters to adapt its behavior to diverse graph topologies, specific signal characteristics, or evolving application requirements, thus limiting its generality.
\end{enumerate}


The advent of the graph fractional Fourier transform (GFRFT) marked a significant step forward by addressing the first limitation. By introducing a fractional order parameter $\alpha$, the GFRFT provides a continuous transition between the vertex domain ($\alpha=0$) and the full GFT domain ($\alpha=1$), enabling analysis in optimal fractional domains. Initially proposed by Wang \textit{et al.} \cite{44}, this transform extends the classical fractional Fourier transform (FRFT) \cite{45,46,47,48,49,50} to graph domains via spectral modulation. Subsequent works have further developed the GFRFT framework in multiple directions \cite{51,52,53,54,55,56}, including a recent hyper-differential operator-based approach \cite{57} that ensures differentiability with respect to $\alpha$, enabling gradient-based learning of the fractional order. This adaptability allows the GFRFT to be integrated into adaptive and task-specific optimization pipelines. Applications such as graph signal denoising and data compression have demonstrated its utility in suppressing noise and preserving structural information \cite{53,58,59,60,61}. Moreover, GFRFT has also been explored in graph signal filtering, its integration into graph neural networks, and a variety of other graph-based learning scenarios \cite{62,63,64,65,66}, underscoring its versatility and establishing it as a valuable tool in the GSP arsenal.

However, while the GFRFT liberates the analysis domain, it does not modify the core transfer function of the GHT. Therefore, the profound limitations of a fixed phase shift and information loss remain unresolved, indicating the necessity for a more comprehensive and generalized framework.

In classical signal processing, the need to address similar challenges leads to the development of the fractional Hilbert transform (FRHT) \cite{67}. The FRHT introduces a powerful dual-parameter framework that decouples domain selection from phase manipulation. An order parameter $\alpha$ controls the rotation in the time-frequency plane, while an angle parameter $\beta$ generalizes the fixed phase shift, allowing for continuously adjustable phase shifts and providing a non-zero, real-valued response ($\cos\beta$) for the direct current (DC) component, thereby completely eliminating information loss. This mature framework has proven highly effective in time-frequency analysis and phase retrieval. Nevertheless, despite its conceptual superiority, this flexible dual-parameter framework has not been systematically formalized for graph signals.

To bridge this gap, this paper proposes a novel {graph FRHT (GFRHT)}, a direct and natural generalization of the classical FRHT to the graph setting. Our main contributions are threefold:
\begin{itemize}
    \item \textbf{Theoretical framework:} We formulate a generalized GFRHT framework, providing its formal definition and foundational interpretations, and we rigorously establish its fundamental properties.
    \item \textbf{Novel graph fractional analytic signal (GFRAS):} We propose the GFRAS, a novel complex-valued signal representation constructed via the GFRHT. The GFRAS serves as a powerful and flexible new tool for advanced feature extraction, enabling high-precision envelope detection, frequency demodulation, and critically, delivering superior performance in downstream tasks such as speech classification.
    \item \textbf{Performance validation:} We experimentally demonstrate the efficacy and superiority of GFRHT over conventional GHT in multiple practical applications, including edge detection, anomalous signal localization, and speech classification.
\end{itemize}

The proposed GFRHT introduces a dual-parameter framework that simultaneously overcomes all aforementioned limitations of the GHT. The fractional order parameter $\alpha$ enables analysis in any fractional domain, while the angle parameter $\beta$ provides continuously tunable phase shifts and preserves information at real eigenvalues. This offers unprecedented flexibility and control for graph signal analysis.

The remainder of this paper is structured as follows. Section \ref{Sec:2} reviews the necessary preliminary concepts. Section \ref{Sec:3} formally defines the proposed GFRHT and elucidates its properties. Section \ref{sec:4} introduces the GFRAS and derives several important concepts, including fractional graph amplitude, phase, and frequency modulation. Section \ref{sec:5} presents experimental results and analyses. Finally, Section \ref{sec:6} concludes the paper with discussions and future directions.

\section{Preliminaries}
\label{Sec:2}

This section reviews the fundamental concepts and mathematical tools that serve as the foundation for the proposed GFRHT. We begin with the GFT, which facilitates spectral analysis of graph signals. Subsequently, we introduce the conventional GHT and its associated concepts for generating GAS. To overcome the limitation of a fixed analysis domain, we then present the GFRFT. Finally, we conclude by recalling the classical FRHT, whose dual-parameter framework provides the direct inspiration for our work.

\subsection{GFT}
\label{subsec:gft}

Consider a graph $\mathcal{G} = (\mathcal{V}, \mathbf{A})$, where $\mathcal{V}$ is the set of $N$ vertices and $\mathbf{A} \in \mathbb{R}^{N \times N}$ is the adjacency matrix. A graph signal is a function $\mathbf{x} : \mathcal{V} \rightarrow \mathbb{R}^N$ that assigns a value to each vertex. The GFT provides a means to analyze $\mathbf{x}$ in the spectral domain defined by the graph structure.

The GFT is defined via the Jordan decomposition of the adjacency matrix
\begin{equation}
\mathbf{A} = \mathbf{U} \mathbf{J} \mathbf{U}^{-1},
\end{equation}
where $\mathbf{U}$ is the matrix of generalized eigenvectors and $\mathbf{J}$ is a block-diagonal Jordan canonical form. The GFT of a signal $\mathbf{x}$ is then given by
\begin{equation}
\hat{\mathbf{x}} = \mathbf{F} \mathbf{x} = \mathbf{U}^{-1} \mathbf{x},
\end{equation}
and the inverse GFT is
\begin{equation}
\mathbf{x} = \mathbf{U} \hat{\mathbf{x}} = \mathbf{F}^{-1} \hat{\mathbf{x}}.
\end{equation}
If $\mathbf{A}$ is diagonalizable, $\mathbf{J}$ reduces to $\mathbf{\Lambda} = \operatorname{diag}(\lambda_1, \lambda_2, \ldots, \lambda_N)$, where $\{\lambda_k\}$ are the eigenvalues of $\mathbf{A}$, and the GFT basis $\mathbf{U}$ becomes orthogonal.

\subsection{GHT and GAS}
\label{subsec:ght}

The GHT is a linear operator that facilitates the construction of an analytic signal on a graph, enabling the extraction of local envelope and phase information.

\subsubsection{GHT}

The GHT of a graph signal $\mathbf{x}$ is defined as a spectral domain operation
\begin{equation}
\mathcal{H}(\mathbf{x}) = \mathbf{F}^{-1} \widehat{\mathbf{H}} \mathbf{F} \mathbf{x},
\label{eq:ght_definition}
\end{equation}
where $\widehat{\mathbf{H}}$ is a diagonal transfer function matrix with elements defined by
\begin{equation}
\widehat{\mathbf{H}}[k,k] = 
\begin{cases}
-j, & \operatorname{Im}(\lambda_k) > 0, \\
0,  & \operatorname{Im}(\lambda_k) = 0, \\
j,  & \operatorname{Im}(\lambda_k) < 0,
\end{cases}
\label{eq:ght_transfer}
\end{equation}
which corresponds to a phase shift of $\mp\tfrac{\pi}{2}$ determined uniquely by the sign of $\operatorname{Im}(\lambda_k)$, 
while annihilating the components associated with real eigenvalues.

\subsubsection{GAS and Related Concepts}

The GAS is constructed from the original signal $\mathbf{x}$ and its GHT
\begin{equation}
\tilde{\mathbf{x}} = \mathbf{x} + j \mathcal{H}(\mathbf{x}).
\label{eq:gas}
\end{equation}
From the GAS, we can derive several key quantities for graph signal analysis.

    {Graph amplitude modulation (GAM):} The GAM is given by the magnitude of the GAS
    \begin{equation}
    \mathcal{A}[k] = |\tilde{\mathbf{x}}[k]|.
    \label{eq:gam}
    \end{equation}
    
    {Graph phase modulation (GPM):} The GPM is given by the phase angle of the GAS
    \begin{equation}
    \phi[k] = \arctan\left( \frac{\operatorname{Im}(\tilde{\mathbf{x}}[k])}{\operatorname{Re}(\tilde{\mathbf{x}}[k])} \right).
    \label{eq:gpm}
    \end{equation}
    
    {Graph frequency modulation (GFM):} The GFM is defined as
    \begin{equation}
    \omega = \phi^{u} - \mathbf{A} \phi^{u},
    \label{eq:gfm}
    \end{equation}
    where $\phi^u$ denotes the unwrapped phase vector of the GAS, which is obtained by performing one-dimensional conventional phase-unwrapping on $\phi$.

\subsection{GFRFT}
\label{subsec:gfrft}

The GFRFT generalizes the GFT by introducing a fractional order parameter $\alpha$ that enables continuous transition between the vertex domain and the spectral domain.

Let the GFT matrix $\mathbf{F}$ have an eigendecomposition $\mathbf{F} = \mathbf{V} \mathbf{\mathbf{D}} \mathbf{V}^{-1}$. The GFRFT matrix of order $\alpha$ is defined as
\begin{equation}
\mathbf{F}^{\alpha} = \mathbf{V} \mathbf{\mathbf{D}}^{\alpha} \mathbf{V}^{-1},
\label{eq:gfrft_definition}
\end{equation}
where $\mathbf{\mathbf{D}}^{\alpha} = \operatorname{diag}(d_1^{\alpha}, d_2^{\alpha}, \ldots, d_N^{\alpha})$. The GFRFT of a signal $\mathbf{x}$ is then
\begin{equation}
\hat{\mathbf{x}}_{\alpha} = \mathbf{F}^{\alpha} \mathbf{x}.
\label{eq:gfrft_operation}
\end{equation}

\noindent{The GFRFT possesses several properties:}
\begin{enumerate}
 \item  \textit{Identity:} $\mathbf{F}^{0} = \mathbf{I}$, which is the identity operator in the vertex domain.
 \item \textit{Completeness:} $\mathbf{F}^{1} = \mathbf{F}$, which is the standard graph Fourier transform operator, mapping to the spectral domain.
 \item \textit{Additivity:} $\mathbf{F}^{\alpha+\beta} = \mathbf{F}^{\alpha} \mathbf{F}^{\beta}$ for any fractional orders $\alpha$ and $\beta$.
 \item \textit{Interpolation:} The transform provides a smooth transition between the vertex and spectral domains as the order $\alpha$ varies.
\end{enumerate}

\subsection{FRHT in Classical Signal Processing}
\label{subsec:frht}

The classical FRHT provides a dual-parameter framework that overcomes the limitations of the conventional Hilbert transform. It serves as the direct inspiration for the graph-based generalization proposed in this work.

For a continuous-time signal $x(t)$, the FRHT is defined as
\begin{equation}
\mathcal{H}_{\alpha, \beta}\{x(t)\} = \mathcal{F}^{-\alpha} \left\{ H_{\beta}(\omega) \cdot \left[ \mathcal{F}^{\alpha}\{x(t)\} \right] \right\},
\label{eq:classical_frht}
\end{equation}
where $\mathcal{F}^{\alpha}$ denotes the FRFT operator, which rotates the signal in the time-frequency plane, and $H_{\beta}(\omega)$ is the fractional transfer function defined as

    \begin{equation}
    H_{\beta}(\omega) = 
    \begin{cases}
    e^{-j\beta}, & \omega > 0, \\
    \cos\beta,   & \omega = 0, \\
    e^{j\beta},  & \omega < 0.
    \end{cases}
    \label{eq:classical_transfer}
    \end{equation}

\noindent The parameters $\alpha$ and $\beta$ provide independent control over two crucial aspects: 
{(i) Fractional order ($\alpha$)}, which specifies the rotation angle of the signal representation 
in the time–frequency plane and determines the analysis domain of the transform; {(ii) Angle ($\beta$)}, which extends the phase shift beyond the fixed $\pm \tfrac{\pi}{2}$ 
and yields a non-zero response at DC ($\omega = 0$).

This dual-parameter framework significantly enhances flexibility in signal manipulation while preventing information loss. The proposed GFRHT in the following section extends this powerful concept to the graph signal domain.

\section{GFRHT}
\label{Sec:3}

\subsection{Definition and Basic Interpretations} 
\label{subsec:def_interpret}

\subsubsection{Mathematical Definition}
\label{subsubsec:math_def}

Formally, the GFRHT of a graph signal $\mathbf{x} \in \mathbb{R}^N$ is defined as a operation parameterized by $\alpha$ and $\beta$

\begin{equation}
\mathcal{H}_{\alpha,\beta}(\mathbf{x}) = \mathbf{F}^{-\alpha} \widehat{\mathbf{H}}_{\beta} \mathbf{F}^{\alpha} \mathbf{x},
\label{eq:GFRHT}
\end{equation}
where $\mathbf{F}^{\alpha}$ denotes the GFRFT matrix, and $\widehat{\mathbf{H}}_{\beta}$ is a diagonal matrix referred to as the {graph fractional transfer function}. The elements of $\widehat{\mathbf{H}}_{\beta}$ are defined as

\begin{equation}
\widehat{\mathbf{H}}_{\beta}[k,k] =
\begin{cases}
e^{-j\beta}, & \operatorname{Im}(\lambda_k) > 0, \\
 \cos\beta,   & \operatorname{Im}(\lambda_k) = 0,\\
e^{j\beta},  & \operatorname{Im}(\lambda_k) < 0,
\end{cases}
\label{eq:graph_transfer_function}
\end{equation}
where $\{\lambda_k\}$ are the eigenvalues of the graph adjacency matrix $\mathbf{A}$.

It is worth noting that when $\beta = 0$ (for any $\alpha \in \mathbb{R}$), the GFRHT reduces to the identity transform, whereas when $\alpha = 1$ and $\beta = \frac{\pi}{2}$, it simplifies to the conventional GHT. This demonstrates that the GFRHT framework unifies and generalizes the traditional GHT, offering more flexible graph signal processing capabilities through the introduction of the fractional-order parameter $\alpha$ and angle parameter $\beta$.

\subsubsection{Fractional Domain Graph Convolution Interpretation} 
\label{subsubsec:conv_interpret}

The GFRHT can be elegantly interpreted as a convolution operation in the fractional domain. Let $\widehat{\mathbf{H}}_{\beta}=\mathrm{diag}(\widehat{\mathbf{h}}_{\beta})$. We can then rewrite Eq. (\ref{eq:GFRHT}) as
\begin{equation}
\mathcal{H}_{\alpha,\beta}(\mathbf{x}) = \mathbf{F}^{-\alpha} \left( (\mathbf{F}^{\alpha} \mathbf{x}) \circ \widehat{\mathbf{h}}_{\beta} \right) = \mathbf{x} \ast_{\alpha} (\mathbf{F}^{-\alpha} \widehat{\mathbf{h}}_{\beta}),
\end{equation}
where $\ast_{\alpha}$ denotes the convolution operator in the $\alpha$-th fractional domain. This reformulation provides a profound insight: the GFRHT of a signal $\mathbf{x}$ is equivalent to filtering it with a kernel $\mathbf{g}_{\alpha,\beta} = \mathbf{F}^{-\alpha} \widehat{\mathbf{h}}_{\beta}$ that is specifically designed in the chosen fractional domain. This establishes a direct link between the proposed transform and the theory of graph filtering, suggesting that the GFRHT can be interpreted as a novel class of graph filters whose characteristics are jointly governed by $\alpha$ and $\beta$.

\subsection{Fundamental Properties} 
\label{subsec:properties}

The GFRHT possesses several mathematical properties that underscore its well-defined behavior and utility. 
\begin{enumerate}
    \item \textit{Periodicity:} The GFRHT is periodic in the angle parameter $\beta$ with period $2\pi$, i.e., $\mathcal{H}_{\alpha,\beta}(\mathbf{x}) = \mathcal{H}_{\alpha,\beta+2\pi}(\mathbf{x})$. This is expected and confirms the circular nature of the phase shift operation.
    \item \textit{Linearity (Superposition):} For any signals $\mathbf{x}, \mathbf{y}$ and scalars $\gamma, \rho \in \mathbb{C}$, $\mathcal{H}_{\alpha,\beta}(\gamma\mathbf{x} + \rho\mathbf{y}) = \gamma\mathcal{H}_{\alpha,\beta}(\mathbf{x}) + \rho\mathcal{H}_{\alpha,\beta}(\mathbf{y})$. This fundamental property confirms it is a linear operator.
    \item \textit{Graph-shift invariance:} Let $\mathbf{A}$ be diagonalizable and $\mathbf{A}_{\alpha} \triangleq \mathbf{F}^{-\alpha} \mathbf{\Lambda} \mathbf{F}^{\alpha}$ be a shift operator in the fractional domain, the GFRHT commutes with the shifted version of the signal in the fractional domain: $\mathcal{H}_{\alpha,\beta}(\mathbf{A}_{\alpha} \mathbf{x}) = \mathbf{A}_{\alpha} \mathcal{H}_{\alpha,\beta}(\mathbf{x})$. This is a desirable trait for linear shift-invariant graph filters, ensuring consistent analysis across the graph.
    \item \textit{Angle addition:} Provided the outputs of the GFRFT (with order $\alpha$) have zero values for real eigenvalues ($\operatorname{Im}(\lambda_k)=0$), the operator satisfies $\mathcal{H}_{\alpha,\beta_1 + \beta_2}(\mathbf{x}) = \mathcal{H}_{\alpha,\beta_2}(\mathcal{H}_{\alpha,\beta_1}(\mathbf{x}))$. This property is foundational for designing cascaded or iterative processing operations.
    \item \textit{Invertibility:} Under the same condition of zero response at real eigenvalues, the transform is invertible: $\mathcal{H}_{\alpha,-\beta}(\mathcal{H}_{\alpha,\beta}(\mathbf{x})) = \mathbf{x}$. This proves that no information is lost during the transformation (barring the real eigenvalue condition), which is a significant improvement over the often non-invertible conventional GHT.
\end{enumerate}
The proofs of these properties are straightforward, so they are omitted here.

\subsection{Computational Aspects and Filtering} 
\label{subsec:comp_filter}

\subsubsection{Direct Computational Implementation}
\label{subsubsec:direct_impl}

The computation of the GFRHT can be efficiently carried out in three steps, leveraging existing fast algorithms for the GFRFT: 

\textbf{Step 1 (Forward transform).} Compute the GFRFT of the graph signal $\mathbf{x}$ with order $\alpha$:  
\begin{equation}
\hat{\mathbf{x}}_{\alpha} = \mathbf{F}^{\alpha} \mathbf{x}.
\end{equation}

\textbf{Step 2 (Spectral filtering).} Multiply the transformed signal by the diagonal mask $\widehat{\mathbf{h}}_{\beta}$:  
\begin{equation}
\hat{\mathbf{y}}_{\alpha} = \hat{\mathbf{x}}_{\alpha} \circ \widehat{\mathbf{h}}_{\beta}.
\end{equation}

\textbf{Step 3 (Inverse transform).} Compute the inverse GFRFT (or the GFRFT with order $-\alpha$) of the filtered signal:  
\begin{equation}
\mathcal{H}_{\alpha,\beta}(\mathbf{x}) = \mathbf{F}^{-\alpha} \hat{\mathbf{y}}_{\alpha}.
\end{equation}

This three-step process—transform, mask, inverse transform—mirrors the classical implementation of filter operations in the frequency domain, ensuring computational clarity and feasibility.

\subsubsection{Polynomial Approximation and Fast Implementation} 
\label{subsubsec:poly_approx}

Furthermore, there exists a profound link between the spectral filter and polynomial filters, which is crucial for practical applications. Let $\mathbf{A}$ be diagonalizable and $\mathbf{A}_{\alpha} \triangleq \mathbf{F}^{-\alpha} \mathbf{\Lambda} \mathbf{F}^{\alpha}$ be a shift operator in the fractional domain, and let $g(x) = \sum_{l=0}^{L-1} h_l x^l$ be a polynomial of order $L-1$. If the filter coefficients $\{h_l\}$ satisfy the system of equations
\begin{equation}
\sum_{l=0}^{L-1} h_l \lambda_k^l = \widehat{\mathbf{H}}_{\beta}[k,k] \quad \text{for } k=1,2,\ldots,N,
\end{equation}
then the GFRHT is equivalent to a linear, shift-invariant polynomial filter in the fractional domain
\begin{equation}
\mathcal{H}_{\alpha,\beta}(\mathbf{x}) = \mathbf{F}^{-\alpha} \widehat{\mathbf{H}}_{\beta} \mathbf{F}^{\alpha} \mathbf{x} = \sum_{l=0}^{L-1} h_l \mathbf{A}_{\alpha}^l \mathbf{x} = g(\mathbf{A}_{\alpha}) \mathbf{x}.
\end{equation}
This equivalence allows the GFRHT operation to be approximated and implemented using fast polynomial approximation techniques, such as those based on Chebyshev polynomials, avoiding the explicit computation of the GFRFT matrix and enabling application to very large-scale graphs. 

\subsubsection{Application to Graph Fractional Spectral Filtering} 
\label{subsubsec:filter_app}



The GFRHT can be naturally viewed as a spectral filter in the fractional domain. 
Consider a noisy observation model $\mathbf{y} = \mathbf{Gx} + \mathbf{n}$, 
where $\mathbf{G}$ is a known transformation and $\mathbf{n}$ denotes additive noise. 
The operation $\mathcal{H}_{\alpha,\beta}(\mathbf{y})$ can be interpreted as a recovered signal obtained through a fractional spectral filter with coefficients $\widehat{\mathbf{h}}_{\beta}[k]$.

A significant advantage of the GFRHT over fixed filters is its adaptability. 
The parameters $(\alpha, \beta)$ can be optimized for specific tasks, rather than being fixed a priori. 
In practice, they may be selected through grid search when the parameter space is small and discrete, 
or by applying gradient-based optimization strategies for efficient learning in continuous domains.  

Formally, the optimal parameters are obtained by solving
\begin{equation}
(\alpha^*, \beta^*) = \arg\min_{\alpha,\beta} \mathbb{E} \left\{ \| \mathcal{H}_{\alpha,\beta}(\mathbf{y}) - \mathbf{x} \|_2^2 \right\},
\label{eq:22}
\end{equation}
which minimizes the mean squared error (MSE) between the recovered signal and the ground truth.


While the optimization framework above presents a compelling application of the GFRHT, the core focus of this paper is to establish its foundational theory and explore its primary role in {analytic signal generation} and {modulation analysis} for feature extraction. Therefore, an empirical study of its performance in denoising, while a valuable direction for future research, falls beyond the scope of the present numerical validation. The simulations that follow in Section \ref{subsec:simulations} are instead designed to systematically characterize the fundamental input-output behavior of the GFRHT transform itself for a clean signal $\mathbf{x}$.

\subsection{Simulation Results of the GFRHT} 
\label{subsec:simulations}
In this section, we present a comprehensive numerical analysis using a directed social network as a case study to demonstrate the practical application and properties of the GFRHT. The network is represented by the adjacency matrix
$$\mathbf{A} = \begin{bmatrix}
0 & 1 & 1 & 0 & 1 \\
0 & 0 & 0 & 1 & 0 \\
0 & 0 & 0 & 1 & 0 \\
0 & 0 & 0 & 0 & 1 \\
1 & 0 & 0 & 0 & 0
\end{bmatrix},$$
where the five nodes correspond to individuals labeled as Alice (node 1), Bob (node 2), Charlie (node 3), David (node 4), and Eve (node 5). The directed edges represent influence relationships within this social network, with a value of 1 in position $(i,j)$ indicating that person $i$ influences person $j$. This asymmetric connectivity pattern results in complex eigenvalues, which is essential for the GFRHT to produce meaningful transformations.

We define a graph signal $\mathbf{x} = [0.8, 0.3, 0.5, 0.2, 0.6]^\top$ on these nodes, representing a measurable attribute such as activity level or opinion strength. To systematically investigate the behavior of the GFRHT, we select discrete values for the parameters $\alpha$ and $\beta$. Specifically, $\alpha$ is set to $0$, $\tfrac{1}{2}$, and $1$, while $\beta$ is set to $0$, $\tfrac{\pi}{4}$, and $\tfrac{\pi}{2}$. The transform is then evaluated for all pairwise combinations of these values.

Fig.~\ref{fig:01} presents the transformation results for all parameter combinations. Each subplot shows the original signal (solid blue line) alongside the transformed signal (dashed vermilion line) for specific $\alpha$ and $\beta$ values. Several key observations emerge from this analysis:

\begin{figure}[htbp]
\centering
\includegraphics[width=\textwidth]{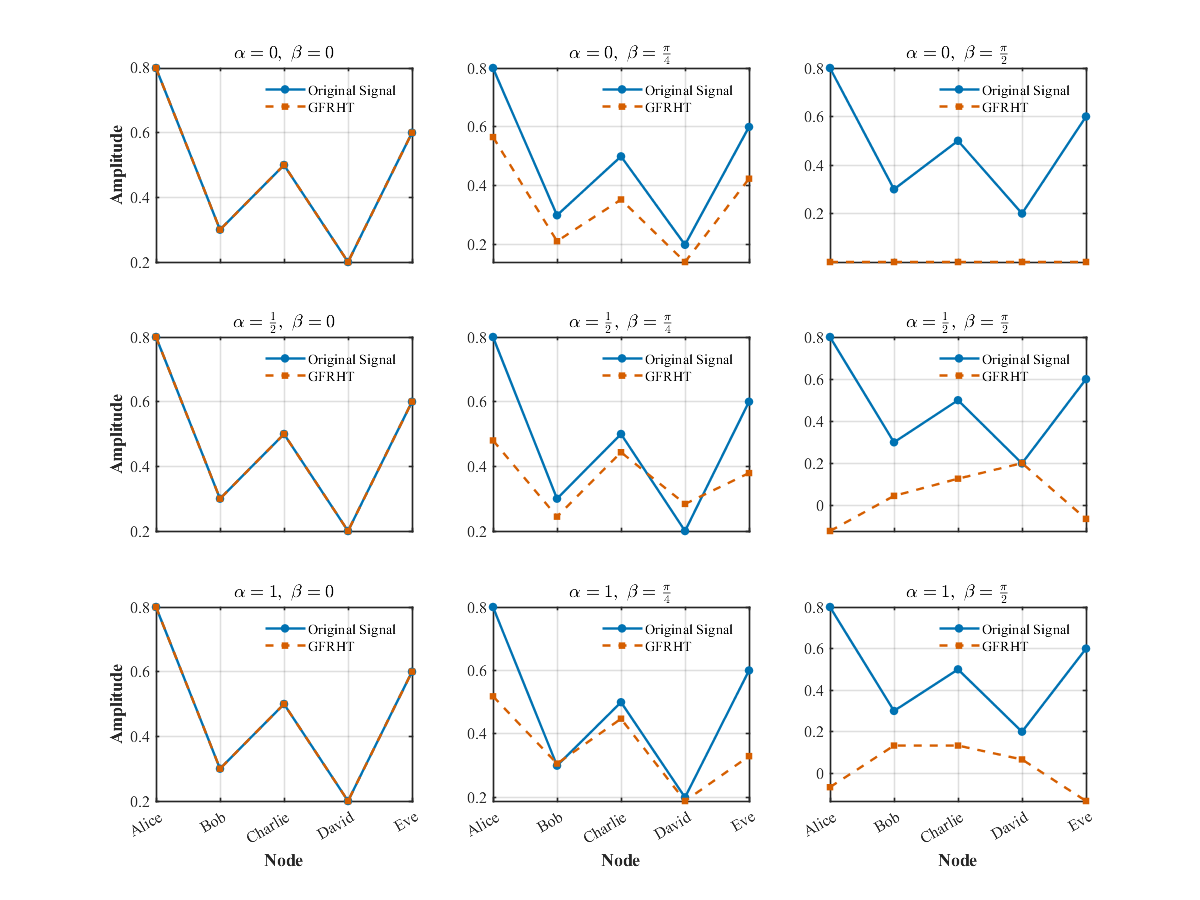}
\caption{Comprehensive parameter study of the GFRHT applied to a social network signal. Each subplot shows the transformation result for specific $\alpha$ and $\beta$ values, with the original signal (blue solid line) and transformed signal (vermilion dashed line). The bottom-right subplot ($\alpha = 1$, $\beta = \tfrac{\pi}{2}$) corresponds to the conventional GHT, providing a connection to established methods. Node labels correspond to individuals in the social network: Alice (1), Bob (2), Charlie (3), David (4), and Eve (5).}
\label{fig:01}
\end{figure}


First, when $\beta = 0$ (i.e., the first column of subplots) for any $\alpha \in \{0, \tfrac{1}{2}, 1\}$, the GFRHT reduces to the identity operation, as expected from the definition of the GFRHT. This serves as a valuable validation of our implementation.

Second, as we increase $\beta$ while keeping $\alpha = 0$, the GFRHT introduces progressively stronger modifications to the signal, demonstrating the role of $\beta$ in controlling the strength of the Hilbert transform component.

Third, the bottom-right subplot ($\alpha = 1$, $\beta = \tfrac{\pi}{2}$) represents a special case where the GFRHT reduces to the conventional GHT. This case is particularly significant as it connects our generalized framework to the well-established GHT, providing a bridge between existing and proposed methodologies.

The systematic variation of both parameters reveals the flexibility of the GFRHT in modulating graph signals. Different parameter combinations emphasize different aspects of the structure of the signal, with some configurations amplifying certain features while suppressing others. This tunability makes the GFRHT a powerful tool for graph signal analysis, particularly in applications where certain frequency components or structural characteristics need to be emphasized or suppressed.

This numerical experiment demonstrates not only the computational feasibility of the GFRHT but also its practical utility in analyzing graph-structured data. The ability to tune both $\alpha$ and $\beta$ parameters provides researchers with a flexible framework for graph signal processing, enabling tailored analyses that can highlight different aspects of graph signals depending on the specific application requirements.

\section{GFRAS}
\label{sec:4}

\subsection{From GAS to GFRAS: A Generalized Analytic Signal}
The conventional GAS, constructed via the GHT, is confined to the spectral domain and suffers from its inherent limitations. The flexibility of the GFRHT provides the necessary framework to generalize this concept. The GFRAS is defined to obtain a complex-valued representation of a real-valued graph signal whose properties—specifically its envelope (amplitude) and phase—can be examined in a proper fractional domain chosen by $\alpha$, and with a phase shift tailored by $\beta$.

\subsection{Definition and Instantaneous Features} \label{subsec:gfras_definition}

Building upon the proposed GFRHT, we define the GFRAS for a real-valued graph signal $\mathbf{x}$ as
\begin{equation}
\tilde{\mathbf{x}}_{\alpha,\beta} = \mathbf{x} + j\mathcal{H}_{\alpha,\beta}(\mathbf{x}) = \mathbf{F}^{-\alpha} (\mathbf{I} + j\widehat{\mathbf{H}}_{\beta}) \mathbf{F}^{\alpha} \mathbf{x}.
\label{eq:GFAS}
\end{equation}

Analogous to the conventional analytic signal, the GFRAS enables the derivation of several fundamental quantities for graph signal analysis, generalizing the concepts of amplitude, phase, and frequency modulation to the fractional-order graph setting. Specifically:  

(i) {Graph fractional amplitude modulation (GFRAM)} is given by the magnitude of the GFRAS,  
\begin{equation}
\mathcal{A}_{\alpha,\beta}[k] = |\tilde{\mathbf{x}}_{\alpha,\beta}[k]| 
= \sqrt{ \left( \operatorname{Re}(\tilde{\mathbf{x}}_{\alpha,\beta}[k]) \right)^2 
+ \left( \operatorname{Im}(\tilde{\mathbf{x}}_{\alpha,\beta}[k]) \right)^2 }.
\label{eq:gfam}
\end{equation}

(ii) {Graph fractional phase modulation (GFRPM)} is defined as the phase angle of the GFRAS,  
\begin{equation}
\phi_{\alpha,\beta}[k] = 
\arctan\left( \frac{\operatorname{Im}(\tilde{\mathbf{x}}_{\alpha,\beta}[k])}
{\operatorname{Re}(\tilde{\mathbf{x}}_{\alpha,\beta}[k])} \right).
\label{eq:gfpm}
\end{equation}

(iii) {Graph fractional frequency modulation (GFRFM)} is given by  
\begin{equation}
\omega_{\alpha,\beta} = \phi_{\alpha,\beta}^{u} - \mathbf{A}\,\phi_{\alpha,\beta}^{u},
\label{eq:gffm}
\end{equation}
where ${\phi}_{\alpha,\beta}^{u}$ denotes the unwrapped phase of the GFRAS. This generalizes the conventional GFM in Eq.~(\ref{eq:gfm}), reducing to it when $\alpha = 1$ and $\beta = \tfrac{\pi}{2}$.

These modulation features provide a powerful tool for analyzing localized spectral behavior and modulations in graph signals within a vertex-frequency framework, effectively generalizing classical modulation concepts from time-frequency analysis to the fractional-order graph setting.

\subsection{Applications of the Extracted Features}
\label{subsec:gfras_applications}

The GFRAS provides a powerful multi-resolution tool for graph signal analysis. The key advantage over the conventional GAS is that by carefully selecting $\alpha$ and $\beta$, one can enhance certain signal features that might be obscured in the canonical domains. The extracted features enable numerous advanced applications:
\begin{enumerate}
    \item {Anomaly detection:} Anomalies often manifest as localized bursts of energy. The GFRAM $\mathcal{A}_{\alpha,\beta}$ derived from a GFRAS, with parameters tuned to maximize the contrast between anomalous and normal regions, can yield a sharper and more localized energy map, significantly improving detection accuracy.
    \item {Modulation analysis and demodulation:} The GFRPM $\phi_{\alpha,\beta}$ can be used to track phase synchrony or the propagation of information waves on the graph. The parameter $\beta$ allows for adjusting the phase response to better align with the underlying modulation process, facilitating tasks like demodulation of graph-bandpass signals.
    \item {Edge detection and feature extraction:} In image processing applications represented on graphs (e.g., pixel grids), the fractional domain parameter $\alpha$ can be tuned to emphasize features at a particular scale. This can lead to more robust edge detection and texture classification across different image types, as the analysis is not locked into a single scale.
\end{enumerate}

The adaptability of the GFRAS through its parameters makes it a versatile tool for a wide range of inference tasks on graphs, moving beyond the capabilities of fixed transforms.

\section{Experiments}
\label{sec:5}
We next illustrate the applications of the proposed GFRHT on both synthesized and real-world signal examples, with a focus on its superior performance in tasks such as edge detection, anomaly detection, and speech signal classification compared to the conventional GHT.

\subsection{Edge Detection Using the GFRHT}
\label{subsec:gfht_anomaly}

\subsubsection{Experimental Setup}

We first investigate the performance of the GFRHT for edge detection in 2D images. The GFRHT, which generalizes the conventional GHT by incorporating two parameters, offers enhanced flexibility in edge detection. To evaluate its effectiveness, we apply the GFRHT to two 40$\times$40 test images: a synthetic image and a gear image.


The graph structure for the image is defined by connecting the $j$-th pixel in the $i$-th row to the $(j+1)$-th pixel in the same row and the $j$-th pixel in the $(i+1)$-th row. The adjacency matrix of this graph is consequently given by the Kronecker product $\mathbf{A} = \mathbf{C} \otimes \mathbf{C}$, where
\begin{equation}
\mathbf{C} = 
\begin{pmatrix}
0 & 1 & 0 & \cdots & 0 \\
0 & 0 & 1 & \cdots & 0 \\
\vdots & \vdots & \vdots & \ddots & \vdots \\
1 & 0 & 0 & \cdots & 0
\end{pmatrix}.
\end{equation}
This graph serves as the shift operator for the subsequent application of the proposed GFRHT. All performance comparisons between the GFRHT and the conventional GHT are conducted under this identical graph construction.

\subsubsection{Results and Analysis}

\begin{figure}[!htp]
	\centering
	\subfloat[]{\includegraphics[width=0.23\textwidth]{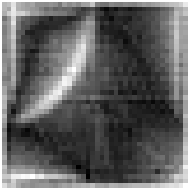}}\hspace{0.02\textwidth}
	\subfloat[]{\includegraphics[width=0.23\textwidth]{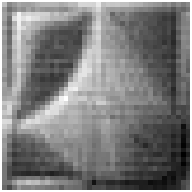}}\hspace{0.02\textwidth}
    \subfloat[]{\includegraphics[width=0.23\textwidth]{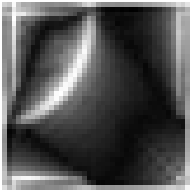}}\\
    \subfloat[]{\includegraphics[width=0.23\textwidth]{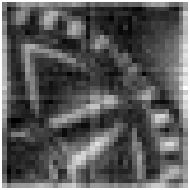}}\hspace{0.02\textwidth}
	\subfloat[]{\includegraphics[width=0.23\textwidth]{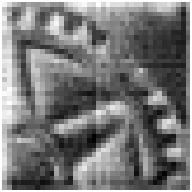}}\hspace{0.02\textwidth}
	\subfloat[]{\includegraphics[width=0.23\textwidth]{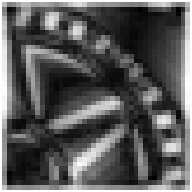}}
    \caption{Edge detection comparison. (a) Synthetic and (b) gear original images. (c) and (d) Results from the GHT. (e) and (f) Results from the proposed GFRHT. The GFRHT provides sharper edges and superior contour continuity, as demonstrated with parameters $(\alpha,\beta) = (\tfrac{7}{10},\frac{\pi}{10})$ for the synthetic image and $(\alpha,\beta) = (\tfrac{4}{5},\tfrac{3\pi}{20})$ for the gear image.}
    \label{fig:1}
\end{figure}

Fig. \ref{fig:1} shows the original images and their respective results after applying the GHT and GFRHT. In both cases, the GFRHT produces sharper, more distinct edges compared to the GHT, particularly in complex regions of the images. This is evidenced by the clearer delineation of boundaries in the gear image, which contains intricate edge structures.

\begin{table}[!htbp]
    \centering
    \caption{Quantitative comparison between GHT and the proposed GFRHT.}
    \begin{tabular}{c c c c c}
        \toprule
        \textbf{Graph Type} & \textbf{Metric} & \textbf{GHT} & \textbf{GFRHT} & \textbf{Improvement} \\
        \midrule
        \multirow{3}{*}{Synthetic Image} & Information Entropy & 0.8424 & \textbf{0.9037} & +7.3\% \\
        & SSIM & 0.1305 & \textbf{0.7949} & +509\% \\
        & Edge Density & 0.2706 & \textbf{0.3194} & +18.0\% \\
        \midrule
        \multirow{3}{*}{Gear Image} & Information Entropy & 0.7909 & \textbf{0.8910} & +12.7\% \\
        & SSIM & 0.0784 & \textbf{0.8372} & +968\% \\
        & Edge Density & 0.2375 & \textbf{0.3081} & +29.7\% \\
        \bottomrule
    \end{tabular}
    \label{tab:1}
\end{table}

Quantitative results further confirm the superior performance of the GFRHT. As shown in Table \ref{tab:1}, we evaluate edge detection performance using standard metrics: information entropy, structural similarity index (SSIM), and edge density. The GFRHT consistently outperforms the GHT in all metrics, with higher information entropy, SSIM values, and edge density, indicating that the GFRHT provides better edge clarity and preserves more detail in the images.

\subsubsection{Discussion}

The experimental results demonstrate that GFRHT is highly effective for edge detection, especially in images exhibiting complex edge structures. The ability of GFRHT to accentuate fine details while preserving sharp boundaries represents a clear advantage over the conventional GHT. This enhanced performance arises from the flexibility afforded by the two tunable parameters \( \alpha \) and \( \beta \), which enable more precise control of the transformation.

The higher values observed in the information entropy, SSIM, and edge density metrics suggest that the GFRHT not only enhances edge detection but also improves the overall image representation, making it a promising tool for applications requiring accurate edge detection and anomaly detection.

However, further investigations are needed to explore the performance of the GFRHT on larger and more diverse datasets. Additionally, the impact of two tunable parameters on edge detection quality should be examined to identify the optimal settings for various types of images.

\subsection{Anomaly Localization under Varying Graph Densities}
\label{subsec:anomaly_detection}

\subsubsection{Experimental Setup}
\label{subsubsec:exp_setup}

We construct a synthetic social network graph comprising 10 communities to validate the effectiveness of the proposed GFRHT in anomaly node localization tasks, with each community containing 6 nodes, resulting in a total of $N = 60$ nodes. The network topology is designed as follows: intra-community edge weights are sampled from a uniform distribution $\mathcal{U}(0,1)$, reflecting strong connectivity within communities, while inter-community edge weights are drawn from $\mathcal{U}(0,0.5)$ and randomly placed between node pairs from different communities, forming a weighted and asymmetric adjacency matrix $\mathbf{A}$. Such graph structures with real edge weights are prevalent in practical applications such as road traffic analysis and brain connectivity studies.

Two distinct graph connectivity densities are considered to demonstrate the potential of our method:
\begin{itemize}
    \item \textbf{Sparse graph}: Inter-community edges constitute {1\%} of the total possible edges in the graph.
    \item \textbf{Dense graph}: Inter-community edges constitute {10\%} of the total possible edges.
\end{itemize}

We normalize the adjacency matrix $\mathbf{A}$ such that its spectral radius $\rho(\mathbf{A}) = 1$. Node labels correspond to the row indices of $\mathbf{A}$. We define a graph signal $\mathbf{x}$ to simulate localized anomalies, where $x_i = 1$ for nodes $i = 18$ to $23$ (located in communities 3 and 4) and $x_i = 0$ elsewhere. We then process this signal using both the conventional GHT and the proposed GFRHT (with parameters $(\alpha,\beta) = (\tfrac{11}{10}, \tfrac{\pi}{20})$) to highlight anomalous nodes in the transform domain.

\subsubsection{Results and Analysis}
\label{subsubsec:results}

For objective performance evaluation, three quantitative metrics are employed: 
{signal-to-noise ratio (SNR, dB)}, {localization precision (Precision@$k$)}, 
and {root MSE (RMSE)}. 

The SNR is defined as 
\begin{equation}
\text{SNR}_{\mathrm{dB}} = 20 \log_{10}\!\left(\frac{\mu_{\text{anomaly}}}{\sigma_{\text{background}}}\right),
\label{eq:SNR}
\end{equation}
where $\mu_{\text{anomaly}}$ denotes the mean response magnitude at the ground-truth anomaly nodes, 
and $\sigma_{\text{background}}$ is the standard deviation of the responses at all non-anomaly nodes. 

The localization precision at top-$k$ nodes (Precision@$k$) is computed as 
\begin{equation}
\text{Precision@}k = \frac{|\mathcal{T} \cap \mathcal{P}_k|}{k},
\end{equation}
where $\mathcal{T}$ is the set of true anomaly indices and $\mathcal{P}_k$ is the set of top-$k$ nodes ranked by the response magnitude. 

The RMSE between the estimated response $\mathbf{y}$ and the ground-truth anomaly indicator $\mathbf{g}$ is given by 
\begin{equation}
\text{RMSE} = \sqrt{\frac{1}{N} \sum_{i=1}^{N} (y_i - g_i)^2}.
\end{equation}
{Table~\ref{tab:2}} summarizes the comparative performance of GHT and GFRHT under both sparse and dense graph configurations.

\begin{table}[!htbp]
\centering
\caption{Comparative performance of GHT and GFRHT for anomaly localization under sparse and dense graph structures.}
\begin{tabular}{@{}lcccc@{}}
\toprule
\textbf{Graph Type} & \textbf{Method} & \textbf{SNR (dB)} & \textbf{Precision@6} & \textbf{RMSE} \\ \midrule
\multirow{2}{*}{Sparse Graph} 
 & GHT   & $11.16$ & $0.667$ & $0.2131$ \\
 & GFRHT & $\mathbf{35.83}$ & $\mathbf{1.000}$ & $\mathbf{0.0294}$ \\ \cmidrule(r){1-1}
\multirow{2}{*}{Dense Graph} 
 & GHT   & $-2.41$ & $0.000$ & $0.3883$ \\
 & GFRHT & $\mathbf{17.47}$ & $\mathbf{1.000}$ & $\mathbf{0.1724}$ \\ \bottomrule
\end{tabular}
\label{tab:2}
\end{table}

The results demonstrate that the proposed GFRHT method {consistently and significantly outperforms} the conventional GHT across all tested scenarios.

In the sparse graph, the GFRHT achieves an SNR of {35.83 dB}, representing a remarkable improvement of {24.67 dB} (a {221.0\%} relative increase) over GHT (11.16 dB), thereby substantially enhancing the distinguishability between the anomalous signal and background noise. Furthermore, the GFRHT attains a perfect {Precision@6} of {1.000}, indicating that the top six nodes with the highest output magnitude perfectly coincide with the true anomalous nodes, whereas GHT correctly identifies only {66.7\%} of them. The RMSE of the GFRHT output is reduced by {86.2\%}, confirming its closer resemblance to the ideal binary anomaly signal.

In the dense graph, the performance of GHT severely deteriorates. Its SNR drops to {-2.41 dB}, implying that the noise power surpasses the signal power, rendering the algorithm virtually ineffective. It fails to identify any true anomalies, resulting in a {Precision@6} of {0.000}. In stark contrast, GFRHT exhibits exceptional robustness, maintaining a respectable SNR of {17.47 dB} and a perfect precision score of {1.000}. Despite the increased signal diffusion due to denser connectivity, GFRHT achieves a {55.6\%} reduction in RMSE, underscoring its efficacy in handling complex connection patterns.

\begin{figure}[!htp]
	\centering
	\subfloat[Sparse Graph]{\includegraphics[width=0.48\textwidth]{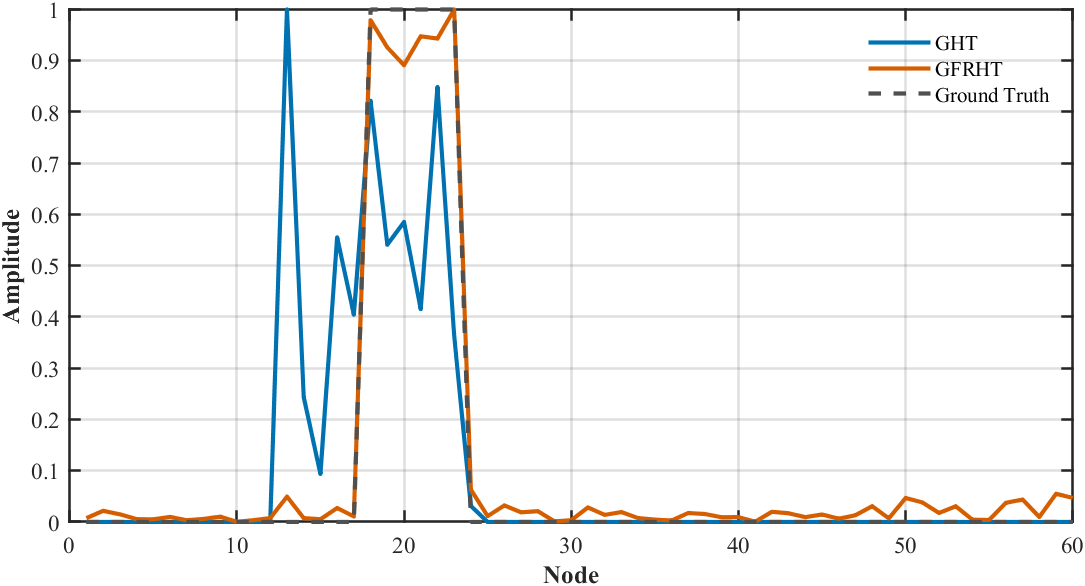}}\hspace{0.02\textwidth}
	\subfloat[Dense Graph]{\includegraphics[width=0.48\textwidth]{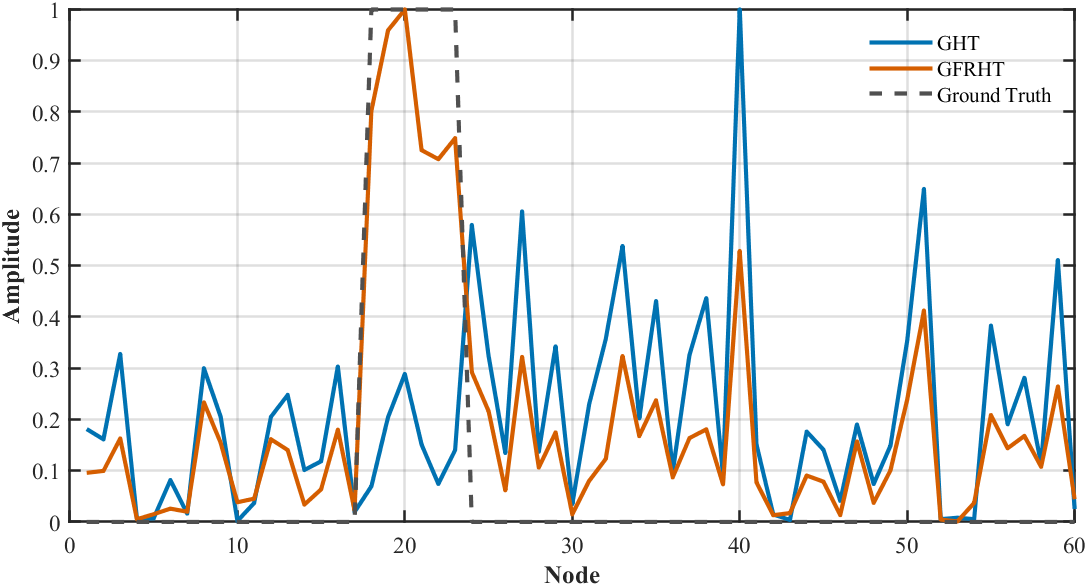}}
	\caption{Visual comparison of anomaly localization performance. (a) and (b) display the output magnitudes of the GHT and the proposed GFRHT for the sparse and dense graph configurations, respectively. The proposed GFRHT produces sharp, concentrated peaks precisely at the true anomaly nodes (18--23) with minimal background artifacts, demonstrating superior precision and noise robustness under both connectivity regimes.}
    \label{fig:2}
\end{figure}

{Fig.~\ref{fig:2}} provides a visual illustration of these findings. For the sparse graph, the output of GFRHT exhibits sharper peaks concentrated around the anomalous region (Nodes 18--23), with minimal background response. For the dense graph, the output of GHT is contaminated by numerous artifacts, making anomaly identification impossible; whereas GFRHT successfully suppresses background noise, clearly highlighting the locations of the anomalous nodes.

In summary, both quantitative and qualitative results confirm that GFRHT not only consistently outperforms GHT across all scenarios but, more importantly, maintains excellent localization performance even in dense graph environments where GHT completely fails. This robust performance firmly establishes the effectiveness of the fractional operator in enhancing the localization properties and resilience of the transform, offering a reliable solution for precise anomaly detection in complex networks.

\subsubsection{Discussion}
\label{subsubsec:discussion}

The experimental results presented in {Section~\ref{subsubsec:results}} unequivocally demonstrate the superior performance of the proposed GFRHT over the conventional GHT for the task of anomaly localization in graph signals. This discussion aims to elucidate the underlying reasons for this performance gap and explore the broader implications of our findings.

The fundamental advantage of GFRHT stems from the additional degrees of freedom provided by the fractional order $\alpha$ and angle $\beta$. While the classical GHT applies a fixed, binary spectral filter ($\pm j$), the fractional framework allows for a more nuanced and continuous modulation of the frequency components of the graph. This flexibility is crucial for adapting to the specific topology of the graph. In the {dense graph} scenario, where excessive inter-community connections cause signal energy to diffuse uncontrollably, the rigid filter of the GHT is unable to isolate the anomaly. The GFRHT, parameterized here with $\alpha= \tfrac{11}{10}, \beta=\tfrac{\pi}{20}$, effectively acts as a customized filter that suppresses this spurious diffusion, thereby concentrating the  energy  of the transform around the true anomalous nodes. This explains the dramatic recovery in SNR from -2.41 dB (GHT) to 17.47 dB (GFRHT) and the leap in precision from 0\% to 100\%.

Furthermore, the perfect {Precision@6} score achieved by GFRHT in both sparse and dense configurations is a particularly significant result. It indicates that the method not only enhances the contrast between signal and noise but also perfectly preserves the {rank ordering} of the anomalous nodes within the output. This property is paramount for practical applications where identifying the exact set of most likely anomalous nodes is required, rather than just a diffuse anomalous region.

The observed performance also highlights the {robustness} of the GFRHT framework. Its exceptional performance across two vastly different graph densities—from a near-block-diagonal structure to a highly connected one—suggests that the method is not overly sensitive to specific topological configurations. This robustness is a desirable property for analyzing real-world networks, which often exhibit complex and varied connection patterns.

\subsection{Detection of Diverse Anomaly Types across Graph Topologies}
\subsubsection{Experimental Setup}

We generate two types of graphs with distinct topological properties: community graphs and scale-free graphs, each containing $N = 50$ nodes. The adjacency matrix $\mathbf{A}$ of each graph is normalized to ensure its spectral radius $\rho(\mathbf{A}) = 1$, which is a standard preprocessing step in graph signal processing.

For each graph type, we introduce three fundamentally different types of anomalies with specific characteristics:

\begin{itemize}
    \item \textbf{Low-frequency anomaly}: The graph signal is constructed by emphasizing the smoothest eigenvector $\mathbf{v}_{\min}$ of $\mathbf{A}$ (associated with the smallest eigenvalue magnitude $|\lambda|$), scaled by a factor of 3. Localized perturbations are then introduced by increasing the amplitude by 1.5 at exactly $k=3$ randomly selected nodes.
    
    \item \textbf{High-frequency anomaly}: The signal is formed by a linear combination of the two most oscillatory eigenvectors ($\mathbf{v}_{\max_1}$ and $\mathbf{v}_{\max_2}$): $\mathbf{x} = \mathbf{v}_{\max_1} + 0.3\mathbf{v}_{\max_2}$. The amplitude is increased by 1.5 at $k=5$ randomly chosen nodes.
    
    \item \textbf{Impulse anomaly}: The signal $\mathbf{x}$ is set to zero at all nodes except for exactly $k=4$ randomly selected nodes, where its value is set to 2.
\end{itemize}



All generated signals are further corrupted by additive white Gaussian noise with a standard deviation of 0.1. Both the conventional GHT with fixed parameters ($\alpha=1, \beta=\tfrac{\pi}{2}$) and the proposed GFRHT are applied to each graph-anomaly combination. For GFRHT, an exhaustive grid search is conducted over the parameter space, with $\alpha \in [0,2]$ sampled at intervals of $0.1$, and $\beta = p \cdot \tfrac{\pi}{2}$ where $p \in [0,2]$ is sampled at intervals of $0.1$. The objective is to determine the optimal parameter configuration by solving
\begin{equation}
(\alpha^\ast, \beta^\ast) = \arg\max_{\alpha,\beta} \ \text{SNR}(\alpha,\beta),
\end{equation}
where SNR is the detection metric defined in Eq. (\ref{eq:SNR}).

\subsubsection{Results and Analysis}

The experimental results unequivocally demonstrate the exceptional performance enhancement achieved by the proposed GFRHT framework compared to the conventional GHT. As quantitatively summarized in Table~\ref{tab:3}, the GFRHT yields extraordinary performance improvements ranging from 361.0\% to an unprecedented 4137.4\% across diverse experimental conditions.

\begin{table}[H]
\centering
\renewcommand{\arraystretch}{0.8}
\caption{Performance comparison between the GHT and GFRHT}
\label{tab:3}
\begin{tabular}{lcccccc}
\toprule
\textbf{Graph Type} & \textbf{Anomaly Type} & $\boldsymbol{\alpha^*}$ & $\boldsymbol{\beta^*}$ & \textbf{GHT SNR} & \textbf{GFRHT SNR} & \textbf{Improvement} \\
\midrule
\multirow{3}{*}{Community} 
  & Low-Frequency  & $\tfrac{7}{5}$   & $\tfrac{9\pi}{10}$  & $1.97$ & $\mathbf{9.08}$  & $+361.0$\%  \\
  & High-Frequency & $\tfrac{7}{5}$   & $\pi$               & $1.36$ & $\mathbf{13.59}$ & $+900.1$\%  \\
  & Impulse        & $\tfrac{1}{10}$  & $\tfrac{19\pi}{20}$ & $0.69$ & $\mathbf{29.12}$ & $+4137.4$\% \\
\midrule
\multirow{3}{*}{Scale-Free} 
  & Low-Frequency  & $\tfrac{11}{10}$ & $\tfrac{9\pi}{20}$  & $0.64$ & $\mathbf{5.22}$  & $+719.8$\%  \\
  & High-Frequency & $\tfrac{1}{5}$   & $\tfrac{\pi}{10}$   & $0.39$ & $\mathbf{12.55}$ & $+3150.6$\% \\
  & Impulse        & $\tfrac{7}{5}$   & $\tfrac{\pi}{20}$   & $1.31$ & $\mathbf{24.84}$ & $+1791.8$\% \\
\bottomrule
\end{tabular}
\end{table}

The optimal fractional parameters exhibit mathematically significant patterns that reflect the intrinsic characteristics of each anomaly type. For low-frequency anomalies, the optimal parameters converge around $\alpha^* = \tfrac{7}{5}$ and $\beta^* = \tfrac{9\pi}{10}$ for community graphs, while scale-free graphs prefer $\alpha^* = \tfrac{11}{10}$ and $\beta^* = \tfrac{9\pi}{20}$. This configuration indicates that moderately smooth fractional transforms combined with specific phase modifications significantly enhance detection of slow-varying anomalies.

For high-frequency anomalies, the optimal parameters demonstrate substantial diversity, with $\alpha^*$ values ranging from $\tfrac{7}{5}$  to $\tfrac{1}{5}$ and $\beta^*$ values spanning from $\pi$  to $\tfrac{\pi}{10}$. This variability underscores the necessity of precise parameter tuning to match the unique spectral characteristics of different graph topologies when detecting oscillatory patterns.

Most remarkably, the GFRHT framework exhibits extraordinary efficacy for impulse anomaly detection, achieving unprecedented improvements of 4137.4\% and 1791.8\% for community and scale-free graphs, respectively. The optimal parameters for impulse anomalies reveal extreme values, with $\alpha^*$ as low as $\tfrac{1}{10}$ and $\beta^*$ approaching $\pi$ ($\tfrac{19\pi}{20}$), indicating that specific fractional orders profoundly enhance the detection of sparse, impulsive disturbances.

\begin{figure}[!htp]
	\centering
	\subfloat[Community graph with low-frequency anomaly]{\includegraphics[width=0.32\textwidth]{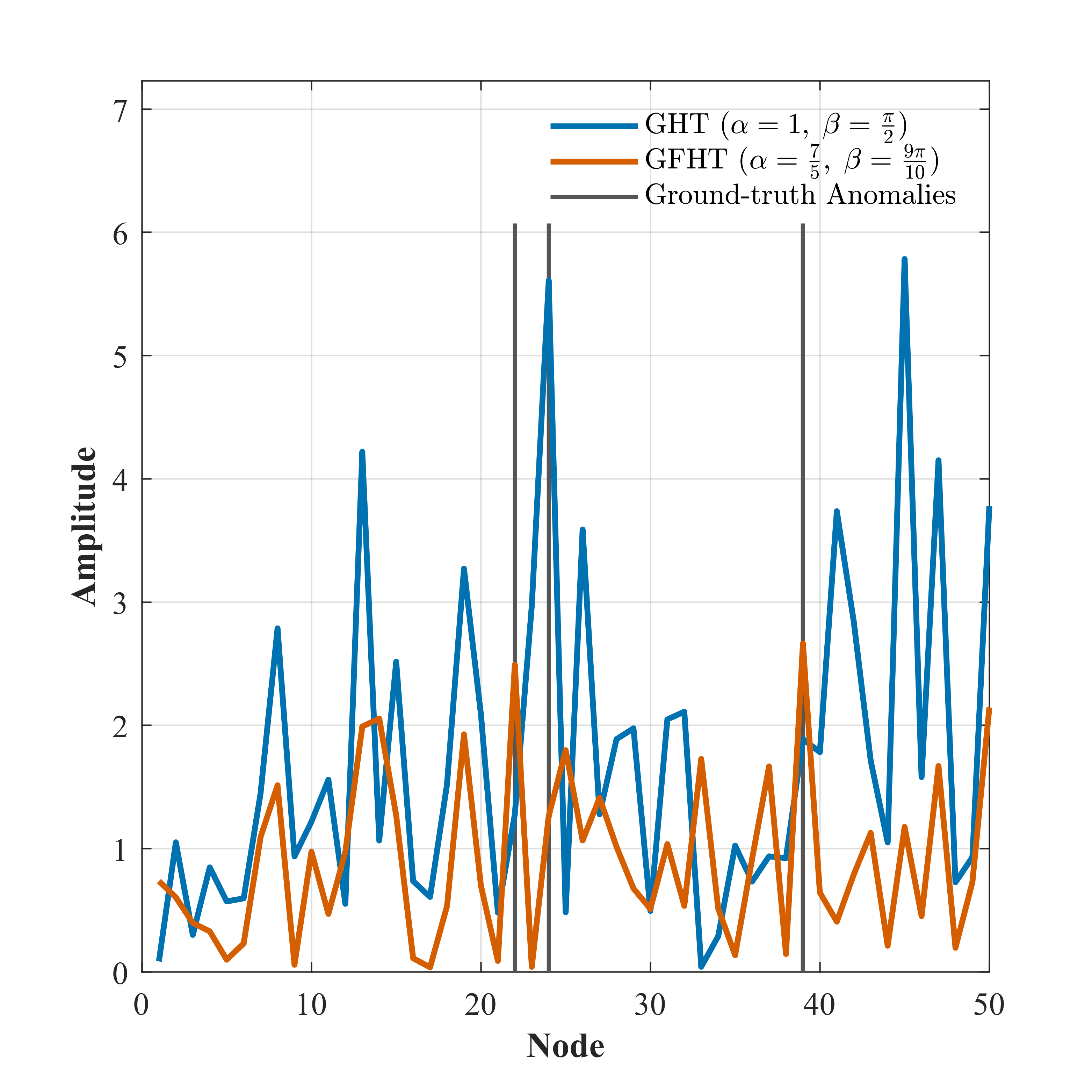}}\hspace{0.001\textwidth}
	\subfloat[Community graph with high-frequency anomaly]{\includegraphics[width=0.32\textwidth]{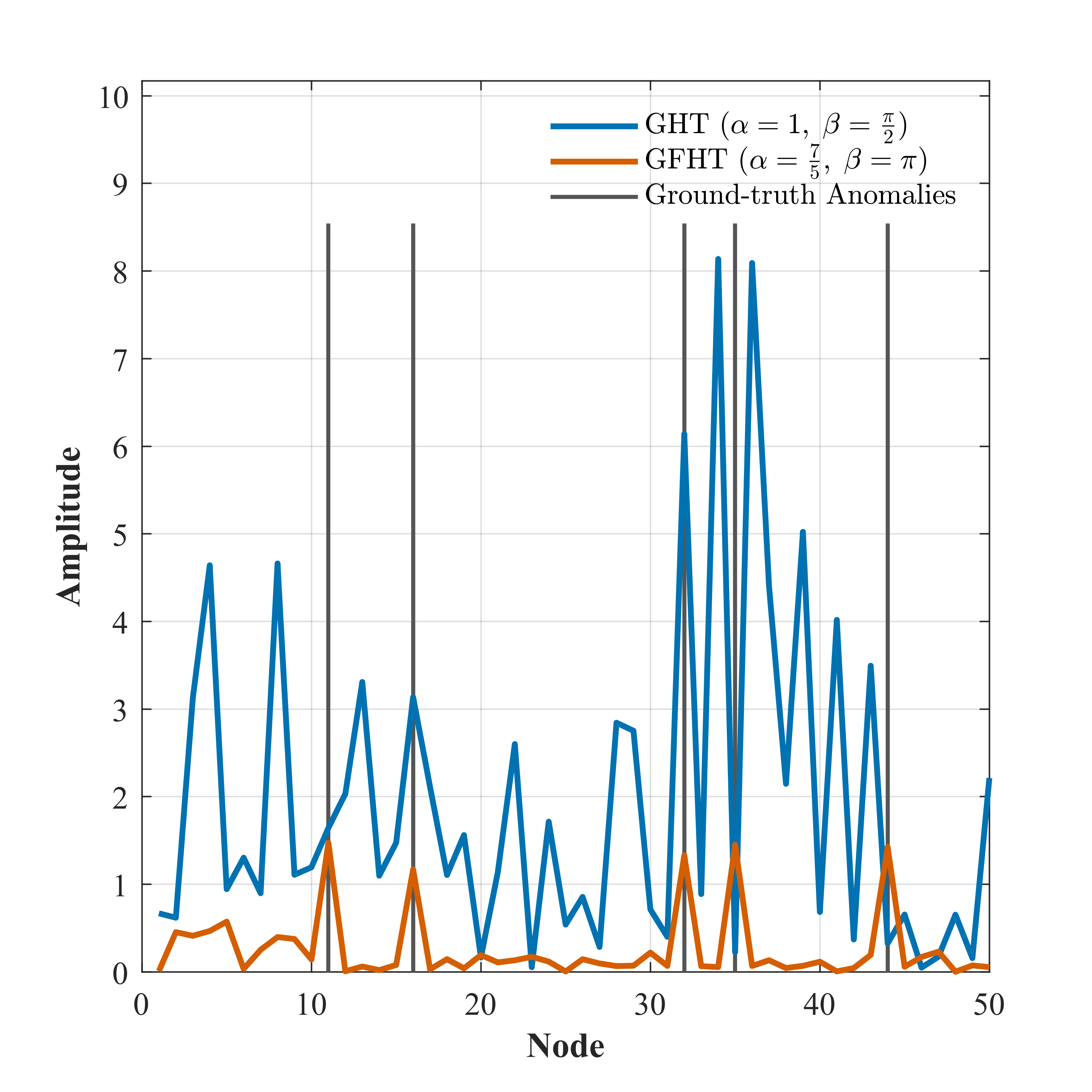}}\hspace{0.001\textwidth}
    \subfloat[Community graph with impulse anomaly]{\includegraphics[width=0.32\textwidth]{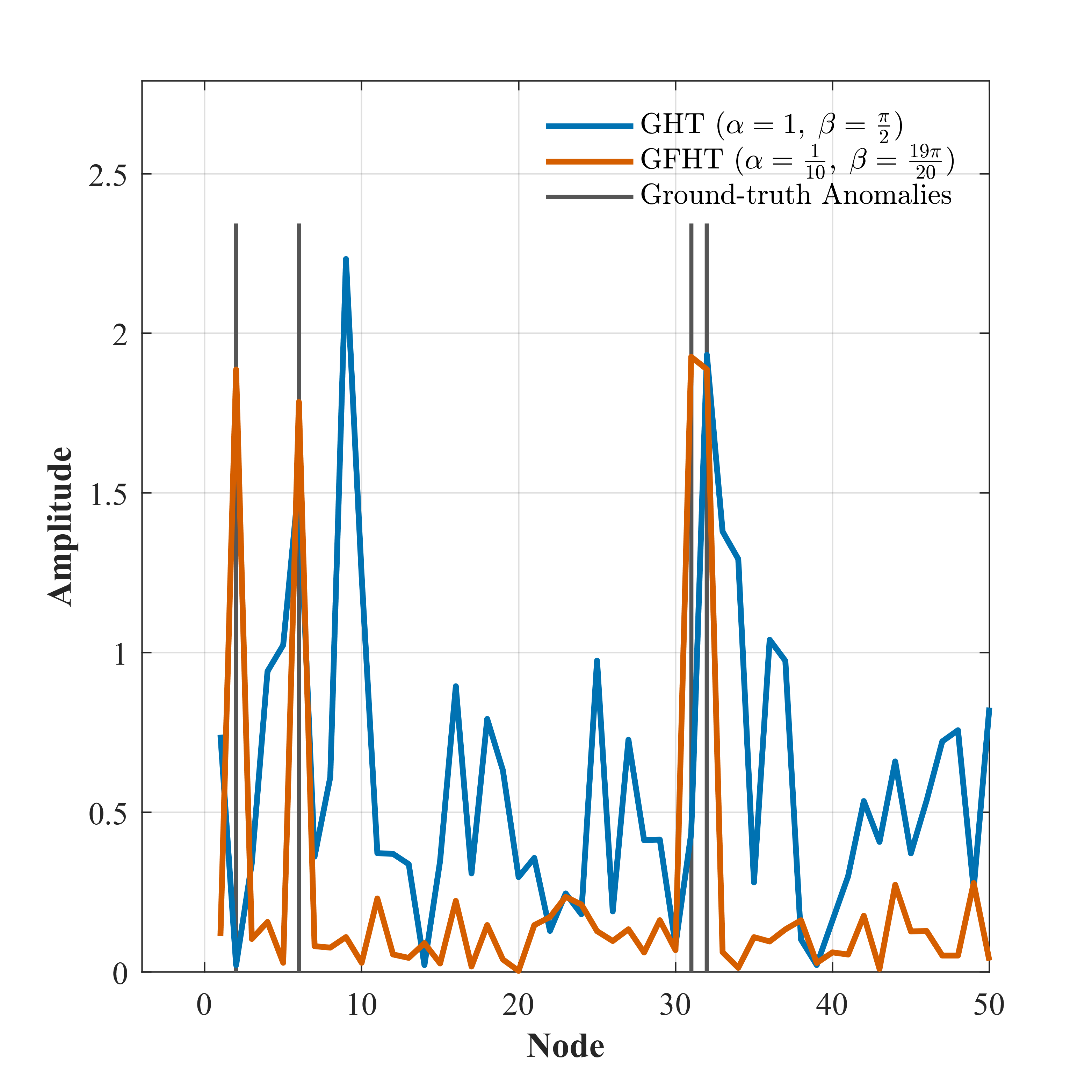}}\\
    \subfloat[Scale-free graph with low-frequency anomaly]{\includegraphics[width=0.32\textwidth]{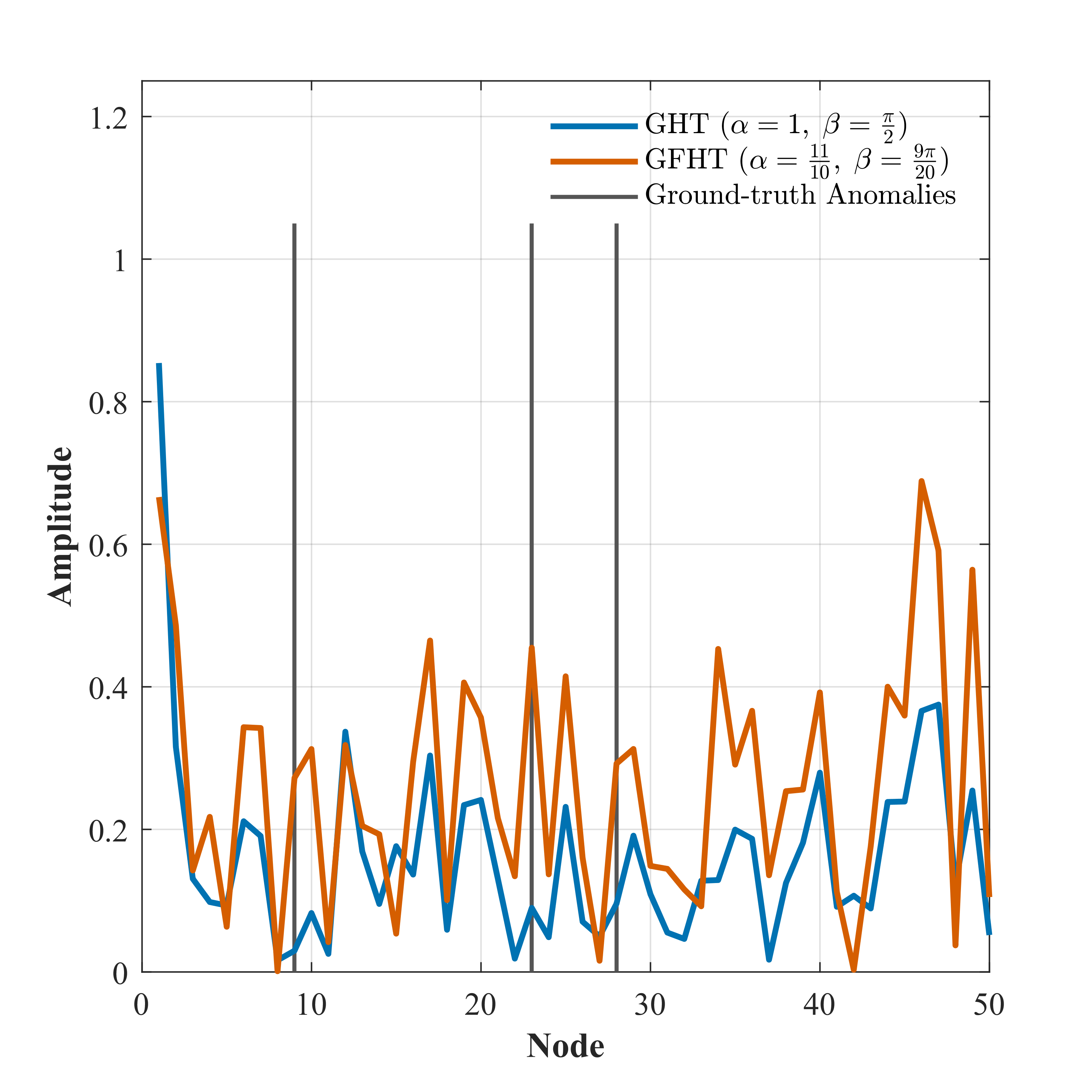}}\hspace{0.001\textwidth}
	\subfloat[Scale-free graph with high-frequency anomaly]{\includegraphics[width=0.32\textwidth]{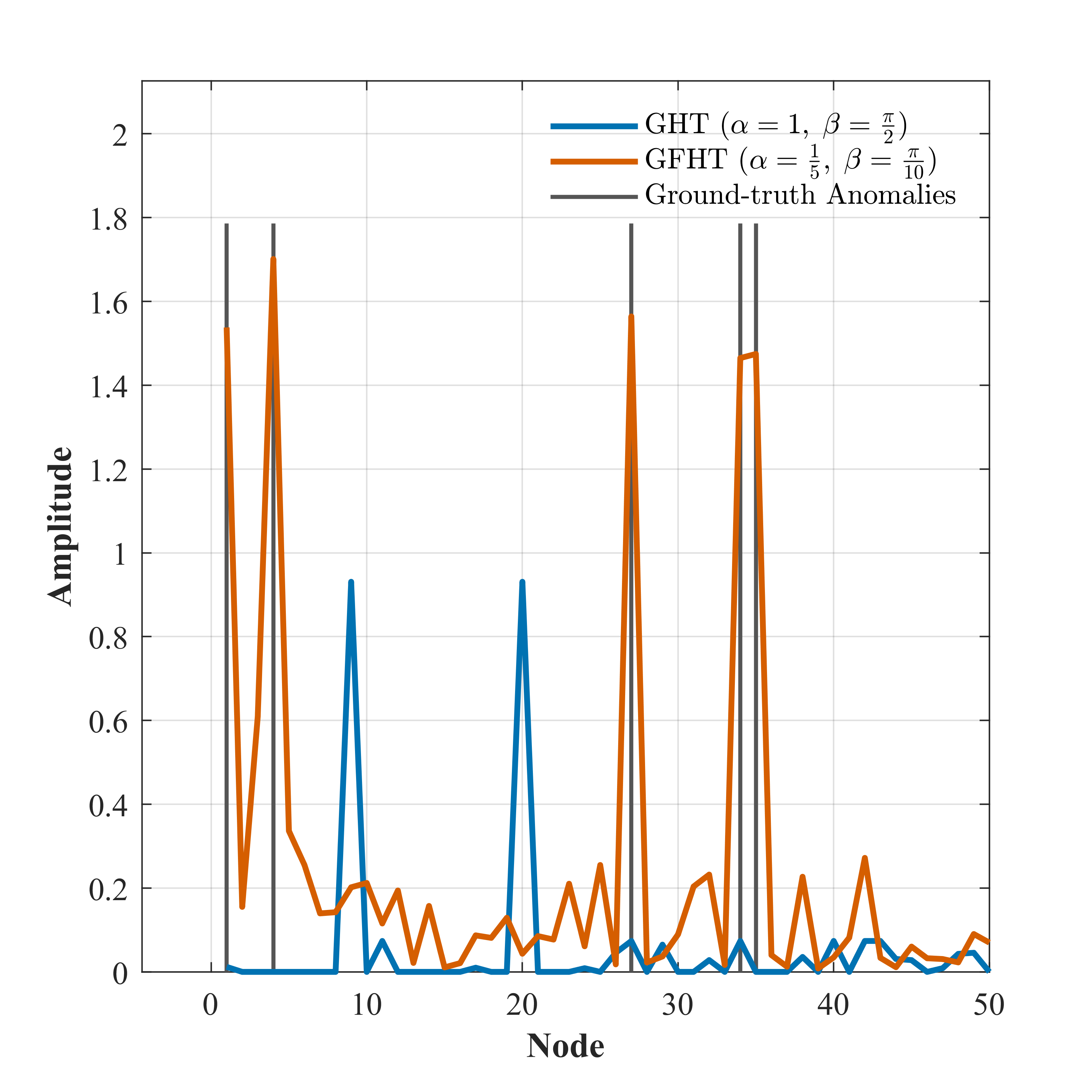}}\hspace{0.001\textwidth}
	\subfloat[Scale-free graph with impulse anomaly]{\includegraphics[width=0.32\textwidth]{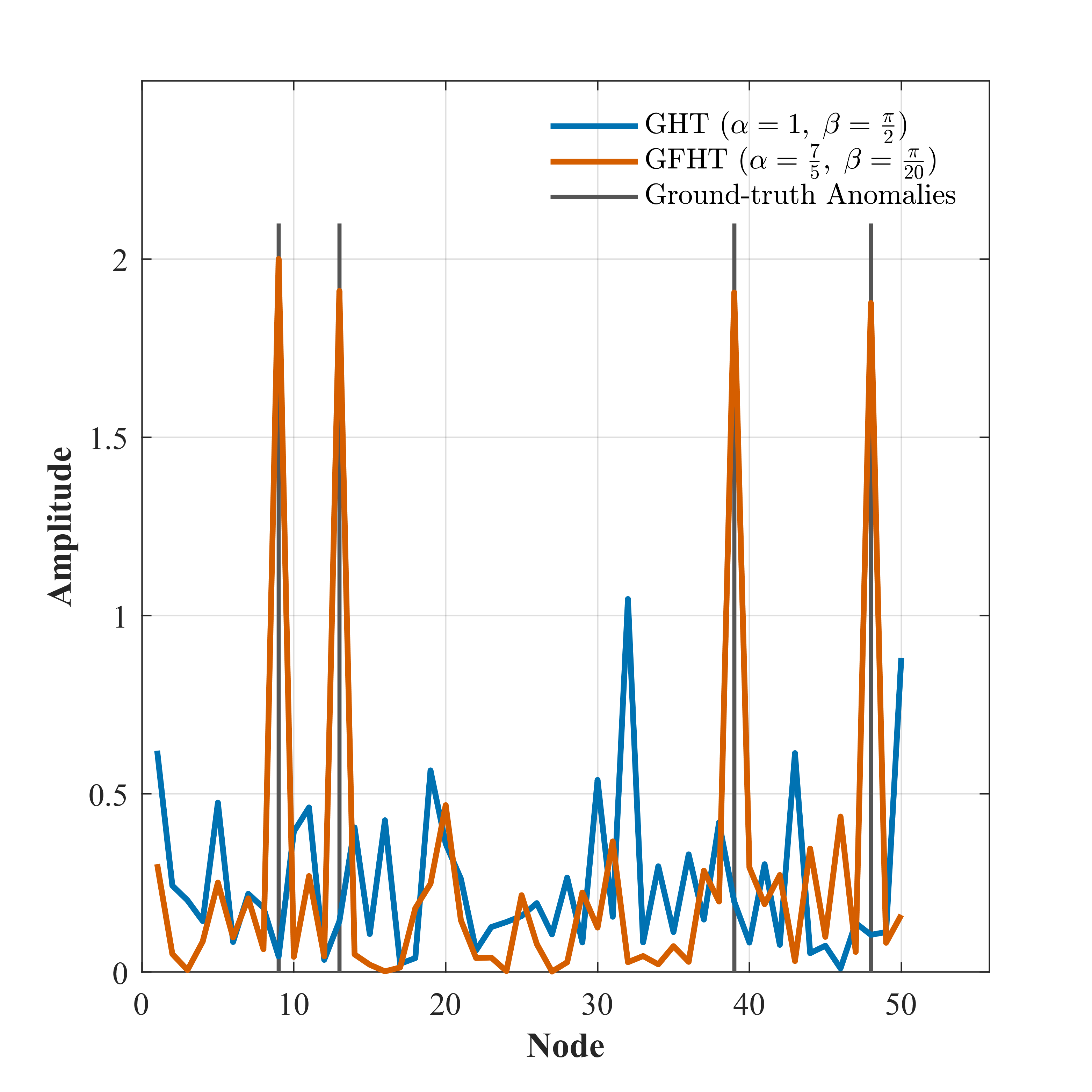}}
	\caption{Visualization of anomaly detection results under different graph topologies and anomaly types. The first row presents results on community graphs, and the second row on scale-free graphs. Each column corresponds to a specific anomaly type: low-frequency (a, d), high-frequency (b, e), and impulse (c, f). The proposed GFRHT shows superior detection capability compared to the baseline GHT.}
    \label{fig:3}
\end{figure}

Fig.~\ref{fig:3} provides comprehensive visual evidence supporting the quantitative findings presented in Table~\ref{tab:3}. The visualization clearly illustrates the superior detection performance of GFRHT across all graph topologies and anomaly types. In each subfigure, the GFRHT response (shown in vermilion) demonstrates significantly sharper and more distinct peaks at the ground-truth anomaly locations compared to the GHT response (shown in blue).

For community graphs with low-frequency anomalies (Fig. \ref{fig:3}(a)), the GFRHT response exhibits enhanced sensitivity to the smooth variations characteristic of this anomaly type. Similarly, for high-frequency anomalies on community graphs (Fig. \ref{fig:3}(b)), the GFRHT effectively captures the oscillatory patterns while suppressing background noise. Most impressively, for impulse anomalies on community graphs (Fig. \ref{fig:3}(c)), the GFRHT produces exceptionally sharp and localized responses precisely at the anomaly locations.

The same trend is observed for scale-free graphs, where the GFRHT consistently outperforms the baseline GHT across all anomaly types (Figs. \ref{fig:3}(d)-(f)). The visual results confirm the remarkable quantitative improvements reported in Table~\ref{tab:3}, particularly the order-of-magnitude enhancements for impulse and high-frequency anomalies.



\subsubsection{Discussion}

The results demonstrate that the GFRHT is highly effective in detecting anomalies, especially for low-frequency, high-frequency, and impulse anomalies. The substantial improvements in SNR indicate that the GFRHT is better at distinguishing anomalies from normal behavior, which is crucial for applications in network anomaly detection, signal processing, and other areas where detecting unusual patterns is critical.

\subsection{Voice Classification Using the GFRAM and GFRFM}
\label{sec:voice_classification}

We extend the application of the proposed GFRHT to a comprehensive voice classification benchmark, aiming to thoroughly validate its effectiveness and generalizability. The primary objective is to investigate whether the GFRAM and GFRFM features, derived from the GFRHT, provide complementary information that enhances classification performance beyond conventional and standard graph-based features.

\subsubsection{Experimental Setup}
\label{sec:voice_setup}

A comprehensive binary classification benchmark is established to evaluate the proposed features. 
The benchmark encompasses four distinct tasks to test generalizability: {Gender Classification} (male vs. female), {Emotion Classification} (happy vs. angry), {Language Classification} (Chinese vs. English), and {Age Group Classification} (minor vs. adult). Speech data are segmented into frames of length $N=50$.

A graph representation of the speech signal is learned from a separate dataset to capture the intrinsic structure of the speech frames. The adjacency matrix $\mathbf{A}^*$ is obtained by solving the following optimization problem, and stability is ensured through appropriate constraints:
\begin{equation}
\mathbf{A}^* = \arg\min_{\mathbf{A}} \|\mathbf{X}_l - \mathbf{A}\mathbf{X}_l\|_2^2 \quad \text{subject to} \quad \operatorname{diag}(\mathbf{A})=\mathbf{0},\ \mathbf{A}\mathbf{1}=\mathbf{1},\ \mathbf{A}^{\mathsf{T}}\mathbf{1}=\mathbf{1}
\label{eq:graph_learning}
\end{equation}
where $\mathbf{X}_l$ is the training data matrix. This learned graph serves as the foundation for all subsequent graph-based feature extraction.

Six different classifiers were compared, each leveraging a unique combination of feature extraction methods:
\begin{itemize}
    \item \textbf{DFT-based:} Uses the magnitudes of the discrete Fourier transform (DFT) coefficients of the conventional 1D AM and FM.
    \item \textbf{GFT-based:} Uses the magnitudes of the GFT coefficients of the GAM and GFM.
    \item \textbf{DFT+GFT:} A concatenation of the feature vectors from the DFT-based and GFT-based classifiers.
    \item \textbf{DFRFT-based:} Uses the magnitudes of the discrete FRFT (DFRFT) coefficients of the 1D AM and FM. The order parameters are treated as learnable parameters and are optimized jointly with the classifier weights via gradient descent.
    \item \textbf{GFRFT-based:} Uses the magnitudes of the GFRFT coefficients of the GFRAM and GFRFM. The fractional order $\alpha$ and angle $\beta$ parameters are treated as learnable parameters and are optimized jointly with the classifier weights via gradient descent.
    \item \textbf{DFRFT+GFRFT:} A concatenation of the feature vectors from the DFRFT-based and GFRFT-based classifiers. This represents our most comprehensive feature set, combining the strengths of fractional transforms in both classical and graph domains.
\end{itemize}


A two-layer neural network is employed as the classifier. The number of neurons in the hidden layer is treated as a tunable hyper-parameter to find a balance between model capacity and computational complexity. The models are trained and tested on data that is strictly held out from the set $\mathbf{X}_l$ used for graph learning. Performance metrics are averaged over multiple runs with random data partitions to ensure statistical significance.

\subsubsection{Results and Analysis}
\label{sec:voice_results}
\noindent \textbf{A. Performance Across Diverse Classification Tasks}

\begin{figure}[!htp]
    \centering
    \includegraphics[width=0.85\linewidth]{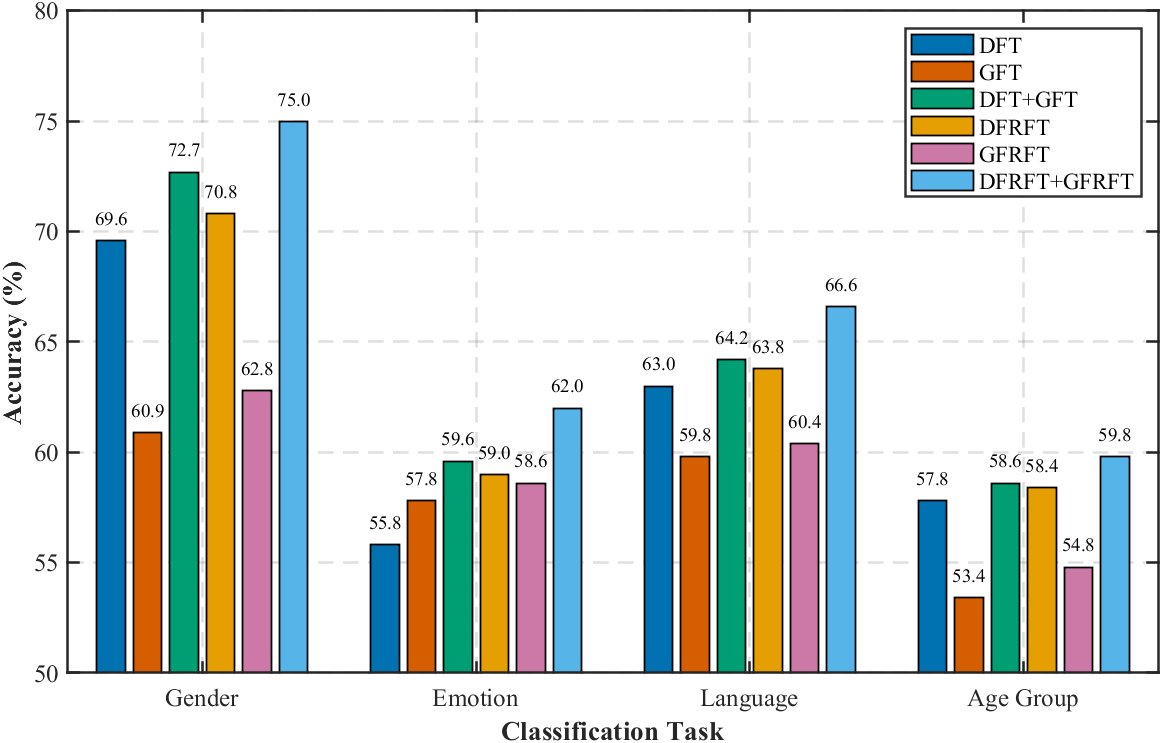}
    \caption{Classification accuracy comparison across different tasks. The proposed DFRFT+GFRFT fusion method consistently achieves the highest performance.}
    \label{fig:4}
\end{figure}

The comprehensive classification accuracies across all four tasks and feature sets are summarized in {Fig.~\ref{fig:4}}. The results lead to several critical observations:
\begin{enumerate}
    \item \textbf{Superiority of fractional graph features:} The proposed {DFRFT+GFRFT} fusion method consistently achieves the highest accuracy across all four classification tasks, demonstrating the universal effectiveness of the fractional features.
    \item \textbf{Significant performance gain:} The DFRFT+GFRFT combination provides a substantial improvement over the previously best-performing DFT+GFT combination, with relative performance gains of {3.16\%}, {4.03\%}, {3.74\%}, and {2.05\%} for gender, emotion, language, and age classification tasks, respectively. This confirms our hypothesis that the fractional transforms capture unique, complementary information.
    \item \textbf{The power of fusion:} The performance hierarchy ({DFRFT+GFRFT} $>$ {DFT+GFT} $>$ {single-method features}) underscores the consistent benefit of feature fusion. Integrating information from both the classical and graph domains, and further enhancing it with fractional orders, yields a more robust and discriminative feature representation.
\end{enumerate}

\noindent \textbf{B. Scalability and Data Efficiency with Increasing Sample Size}

\begin{figure}[tbp]
    \centering
    \includegraphics[width=0.75\linewidth]{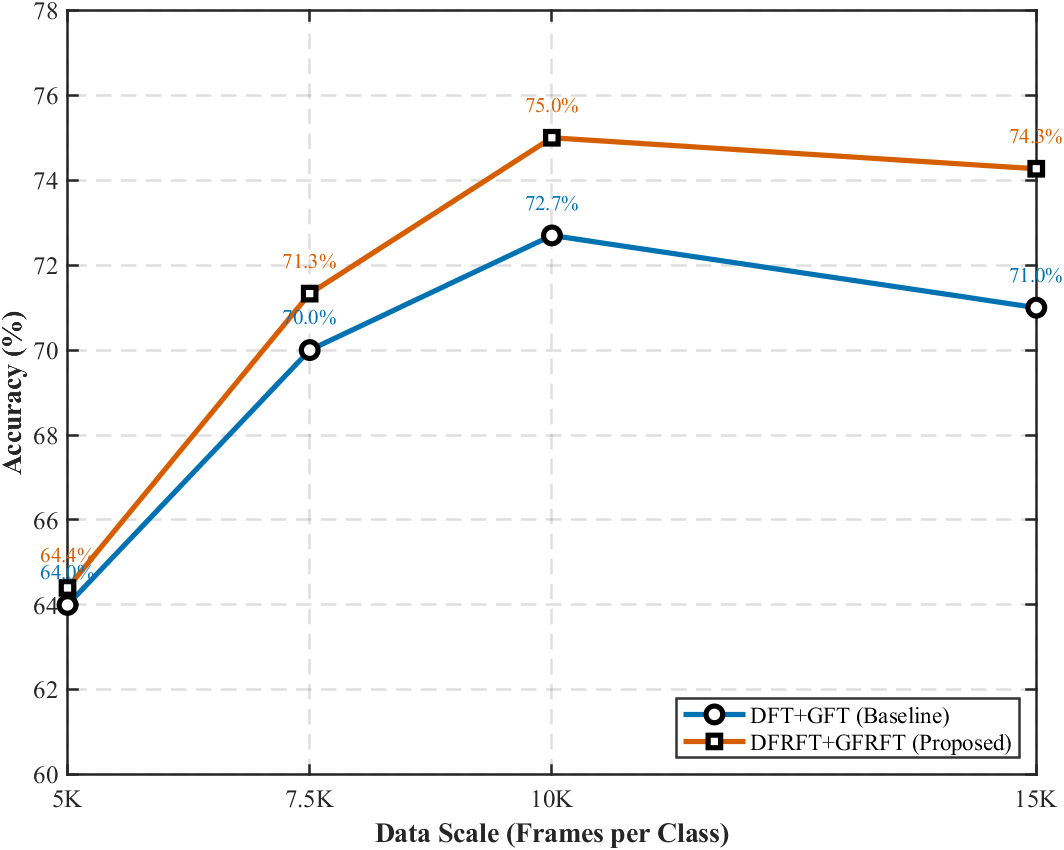}
    \caption{Learning curve showing performance versus data scale for the gender classification task.}
    \label{fig:5}
\end{figure}

To investigate the scalability of the proposed method, we evaluate its performance on the gender classification task with increasing amounts of training data. The results, detailed in {Table~\ref{tab:4}} and {Fig.~\ref{fig:5}}, reveal a crucial trend: the performance advantage of the proposed DFRFT+GFRFT fusion method is maintained across data scales.

\begin{table}[!htbp]
\centering
\caption{Classification accuracy (\%) for the gender task under different data scales.}
\label{tab:4}
\begin{tabular}{@{}ccccccc@{}}
\toprule
\textbf{Data Scale (frames)} & \textbf{DFT} & \textbf{GFT} & \textbf{DFT+GFT} & \textbf{DFRFT} & \textbf{GFRFT} & \textbf{DFRFT+GFRFT} \\
\midrule
5,000       & 62.20        & 53.20        & 64.00            & 63.00          & 58.40          & \textbf{64.40}       \\
7,500       & 65.47        & 59.07        & 70.00            & 67.07          & 59.87          & \textbf{71.33}       \\
10,000     & 69.60        & 60.90        & 72.70            & 70.80          & 62.80          & \textbf{75.00}       \\
15,000      & 68.80        & 61.87        & 71.00            & 70.53          & 62.20          & \textbf{74.27}       \\
\bottomrule
\end{tabular}
\end{table}

This analysis yields two key insights: (i) {Consistent superiority:} The DFRFT+GFRFT fusion strategy consistently outperforms all other methods at every data scale, demonstrating its robustness and reliability; (ii) {Performance trend:} The accuracy of both the baseline and proposed methods shows a strong positive correlation with the amount of data from 5,000 to 10,000 frames, achieving peak performance at the 10,000-frame scale. The observed slight performance degradation at the 15,000-frame mark for both methods is analyzed in the Discussion section.


\subsubsection{Discussion}
\label{sec:voice_discussion}

The experimental results from both diverse tasks and varying data scales unequivocally demonstrate that the features derived from the proposed GFRHT framework provide a significant and robust advantage for voice classification.

The superior performance of the DFRFT+GFRFT classifier establishes that:
\begin{itemize}
    \item \textbf{Fractional transforms offer a richer representation:} By enabling analysis in an optimal fractional domain, the DFRFT and GFRFT uncover subtle, discriminative features in the speech signal that are lost in the conventional Fourier or graph Fourier domains.
    \item \textbf{Synergy between domains:} The classical (DFRFT) and graph-based (GFRFT) fractional transforms are not redundant; they capture complementary aspects of the structure of the signal. Their fusion creates a more powerful feature set than any single method can provide.
    \item \textbf{Data-effective and robust:} The method not only works well on limited data but also maintains its advantage across different data scales, making it suitable for real-world applications.
\end{itemize}



\textbf{Analysis of Performance at Largest Data Scale:}
The observed performance plateau and slight decrease at the largest data scale (15,000 frames) for both methods is attributed to the fixed capacity of the neural network classifier, which is kept constant across all data scales. While sufficient for smaller datasets, this fixed capacity becomes a limiting bottleneck for larger datasets, preventing further learning. Crucially, the performance gap between the proposed and baseline methods persists even at this scale. This indicates that the quality of the features extracted by DFRFT+GFRFT remains superior, and the performance ceiling is imposed by the classifier itself, not the feature extraction front-end.

\section{Discussions and conclusions}
\label{sec:6}
\subsection{Discussion}

This paper introduces the GFRHT, a novel dual-parameter generalization of the conventional GHT. The GFRHT successfully overcomes the fundamental limitations of the GHT by enabling analysis in an arbitrary fractional domain via the order parameter $\alpha$, while the angle parameter $\beta$ allows for continuous phase shift adjustment and eliminates information loss at real eigenvalues. This flexibility is achieved without sacrificing interpretability or computational feasibility, as demonstrated through rigorous theoretical analysis and comprehensive experiments.

The experimental results consistently validate the superiority of the GFRHT over the GHT across a diverse range of applications. In edge detection, the GFRHT produced sharper and more continuous edges, particularly in complex image regions, by emphasizing features at optimal fractional scales. For anomaly detection, the GFRHT exhibited remarkable robustness across varying graph densities and anomaly types, significantly outperforming the GHT in both sparse and dense graphs. The method’s ability to preserve the rank ordering of anomalous nodes—achieving perfect precision in many cases—highlights its practical utility in real-world scenarios where exact localization is critical. In speech classification, the fusion of features derived from the GFRHT with those from the classical DFRFT yielded state-of-the-art performance across diverse tasks. The consistent performance gains across data scales further attest to the robustness and generalizability of the proposed framework, underscoring the complementary and enriched representation provided by fractional analysis in both classical and graph domains.

The superior performance of the GFRHT stems from its generalized dual-parameter framework. The fractional order $\alpha$ allows the transform to operate in an optimal domain that best represents the signal characteristics, moving beyond the rigid vertex or spectral domains. Simultaneously, the angle parameter $\beta$ provides a continuous phase shift, replacing the  binary $\pm j$ operation of the GHT, and crucially preserves energy at real eigenvalues via the $\cos\beta$ term. This adaptability allows the GFRHT to function as a custom-designed filter tailored to the specific graph topology and signal content, a capability the fixed GHT fundamentally lacks.

However, two main limitations warrant attention. First, the computational complexity of the GFRHT is significantly higher than that of the parameter-free GHT. Specifically, the GHT relies on a single eigen-decomposition of the graph adjacency matrix, whereas the GFRHT requires two eigen-decompositions, which substantially increases its computational cost. While the established link to polynomial filters offers a clear path for efficient computation, this overhead remains a practical consideration for large-scale applications. Second, the current validation, while comprehensive, primarily relied on synthetic and structured real-world data. Further testing on more complex, noisy, and dynamic real-world networks (e.g., time-varying graphs or multi-layer networks) is necessary to fully establish the generalizability and resilience of the method.

\subsection{Conclusion}

In conclusion, we propose and formalize the GFRHT, a flexible and powerful tool for graph signal processing that generalizes the conventional GHT through a principled dual-parameter framework. By enabling continuous interpolation between domains and tunable phase shifts, the GFRHT addresses the core limitations of the GHT and offers unprecedented control over graph signal analysis.

Theoretical properties such as linearity, shift-invariance, and invertibility provide a rigorous mathematical foundation for the GFRHT. Empirical evaluations in edge detection, anomaly localization, and speech classification further verify its effectiveness, revealing significant and consistent performance improvements over the conventional GHT. With its adaptability and robustness, the proposed GFRHT emerges as a promising tool for a broad spectrum of graph-based inference tasks, encompassing computer vision, network analysis, and biomedical signal processing.

Future work will focus on several key areas that directly address the current limitations and explore new frontiers: (i) Extending the GFRHT to dynamic graph settings to handle temporally evolving structures, thus validating its performance on the time-varying networks discussed above; (ii) To mitigate the identified computational overhead, we will leverage the polynomial filter interpretation and modern hardware acceleration for large-scale graphs; 3) Exploring the integration of the GFRHT with deep learning frameworks for end-to-end feature learning and analysis on graphs. Additionally, we plan to investigate its applicability in other domains (e.g., bioacoustics, neuroimaging, social network analysis) to thoroughly assess its generalizability across the complex and noisy real-world scenarios outlined in our discussion.

The GFRHT represents a significant advancement in graph signal processing, providing a unified, flexible, and high-performance framework for the analysis of complex signals on graphs.



\bibliographystyle{elsarticle-num-names}
\bibliography{reference}
\biboptions{numbers,sort&compress}





\end{sloppypar}
\end{document}